\documentclass[11pt]{article}
\usepackage{jheppub}
\usepackage[utf8]{inputenc}
\usepackage[T1]{fontenc}  
\usepackage{epstopdf}
\usepackage{enumitem}
\usepackage{tcolorbox}

\usepackage{jheppub}
\usepackage{mathrsfs}
\usepackage{psfrag}
\usepackage{color}
\usepackage{hyperref}

\usepackage{slashed}
\usepackage{feynmp-auto}
\usepackage{simplewick}
\usepackage{cancel}

\usepackage{todonotes}

\usepackage{amsmath}
\usepackage{amsfonts}
\usepackage{graphicx}
\usepackage{amssymb}
\usepackage{xcolor}

\usepackage{mathtools}

\usepackage{amsmath,bbm,array,amsfonts,graphicx,wrapfig,arydshln,lscape,float,multirow,longtable,rotating,makecell}
\usepackage{url}

\DeclareMathAlphabet\mathbfcal{OMS}{cmsy}{b}{n}

\newcommand{\fr}{\frac}

\newcommand{\ra}{\rangle}

\newcommand{\p}{\partial}
\newcommand{\ve}{\varepsilon}
\newcommand{\la}{\langle}

\newcommand{\eq}{\begin{equation}}
\newcommand{\eqe}{\end{equation}}
\newcommand{\eqa}{\begin{eqnarray}}
\newcommand{\eqae}{\end{eqnarray}}
\newcommand{\nn}{\nonumber}
\newcommand{\bn}{\begin{enumerate}}
\newcommand{\en}{\end{enumerate}}
\def\beq#1\eeq{\begin{align}#1\end{align}}

\newcommand{\eqc}[1]{(\ref{#1})}

\def\CO{{\mathcal O}}

\def\CA{{\mathcal A}}

\def\CC{{\mathcal C}}

\def\CE{{\mathcal E}}

\def\CM{{\mathcal M}}

\def\CO{{\mathcal O}}

\def\CR{{\mathcal R}}
\def\CS{{\mathcal S}}

\def\ve{\varepsilon}
\def\vth{\vartheta}

\def\vph{\varphi}
\def\vf{\varphi}

\def\half{\frac{1}{2}}

\def\OC{{\rm oc}}
\def\COM{{\rm com}}

\def\TN{{\rm TN}}
\def\eff{{\rm eff}}

\newcommand{\bfig}{\begin{figure}}
\newcommand{\efig}{\end{figure}}

\def\abs#1{{\left| #1 \right|}}

\def\bl#1\el{\begin{align} #1 \end{align}}
\def\bg#1\eg{\begin{gather} #1 \end{gather}}
\newcommand{\fig}[1]{figure \ref{#1}}
\def\bld#1\eld{\begin{aligned} #1 \end{aligned}}
\def\bgd#1\egd{\begin{gathered} #1 \end{gathered}}

\newcommand{\bra}[1]{\langle{#1}|}
\newcommand{\ket}[1]{|{#1}\rangle}

\newcommand{\sbra}[1]{ [{#1} |}
\newcommand{\sket}[1]{ | {#1} ]}

\renewcommand{\tt}{\texttt}

\newcommand{\RN}[1]{%
  \textup{\uppercase\expandafter{\romannumeral#1}}%
}

\newcommand{\be}{\begin{equation}}
\newcommand{\ee}{\end{equation}}
\newcommand{\ba}{\begin{align}}
\newcommand{\ea}{\end{align}}
\newcommand{\bi}{\begin{itemize}}
\newcommand{\ei}{\end{itemize}}

\let\a=\alpha \let\b=\beta \let\g=\gamma  \let\e=\epsilon
  \let\th=\theta  \let\k=\kappa
\let\l=\lambda \let\m=\mu \let\n=\nu  \let\p=\pi  
\let\s=\sigma \let\t=\tau  \let\f=\phi  
   \let\G=\Gamma \let\D=\Delta \let\Th=\Theta \let\L=\Lambda
    \let\F=\Phi 
 \let\W=\Omega

\let\ph=\phantom
\let\pt=\partial

\makeatletter
\newcommand*{\Rom}[1]{\expandafter\@slowromancap\romannumeral #1@}
\makeatother

\makeatletter
\newcommand*{\rom}[1]{\expandafter\romannumeral #1}
\makeatother

\let\nn=\nonumber

\def\beq#1\eeq{\begin{align}#1\end{align}}

\title{Gravitational Dyonic Amplitude at One-Loop and its Inconsistency with the Classical Impulse}
\author[1,2]{Jung-Wook Kim}
\author[3]{Myungbo Shim}

\affiliation[1]{Centre for Research in String Theory, School of Physics and Astronomy,\\Queen Mary University of London, Mile End Road, London E1 4NS, United Kingdom}
\affiliation[2]{Department of Physics and Astronomy, Seoul National University, Seoul 08826, Korea}
\affiliation[3]{Department of Physics and Research Institute of Basic Science, Kyung Hee University,\\Seoul 02447, Korea}

\emailAdd{jung-wook.kim@qmul.ac.uk}
\emailAdd{mbshim1213@khu.ac.kr}

\abstract{The recent proposal~\cite{Huang:2019cja,Emond:2020lwi} of implementing electric-magnetic duality rotation at the level of perturbative scattering amplitudes and its generalisation to gravitational context where usual gravitational mass is rotated to the NUT parameter of the Taub-NUT spacetime opens up an interesting avenue for studying NUT-charged objects as dynamical entities, in contrast to the usual approach where NUT-charged objects are considered as a static background. We extend the tree-order analysis to one-loop order, and find a disagreement between geodesic motion on Taub-NUT background and impulse computation of scattering amplitudes. As a by-product of our analysis, we find a relation between tidal response parameters and resonance excitation parameters in the language of quantum field theory scattering amplitudes.}

\begin{document}
\begin{flushright}
\vspace{10pt} \hfill{QMUL-PH-20-30} \vspace{20mm}
\end{flushright}
\maketitle

\section{Introduction}

During the expedition for a solution of Einstein's equation having angular momentum, Taub-NUT spacetime had been discovered independently by Taub and Newman, Unti, and Tamburino~\cite{Taub:1950ez,Newman:1963yy}. The solution has an additional parameter mistaken for angular momentum at the time of discovery, the so-called NUT charge, and the spacetime can be considered as a simple generalisation of the Schwarzschild solution~\cite{Misner:1963fr}. Interestingly, Taub-NUT solutions cannot be extrapolated or constructed from summing over higher order corrections of post-Newtonian expansion or modified gravitational theories \cite{Ramaswamy81, Ramaswamy:1986kf,Kol:2020zth}, but it is one of the simplest, Hamilton-Jacobi soluble \cite{Carter:1968ks}, and exact solutions of the vacuum Einstein equations. Although the spacetime with angular momentum has later been identified as the Kerr solution, the physical meaning and effects of NUT charges has still been studied, and the NUT parameter has later been identified as a source of locally rotating frame or gravitomagnetic force which becomes the Coriolis force in (post)-Newtonian regime~\cite{bonnor69,Dowker74}, being the charge of dual supertranslation symmetry~\cite{Huang:2019cja,Alawadhi:2019urr,Kol:2019nkc}. It was also found that angular momentum of Kerr spacetime and NUT parameter of Taub-NUT spacetime can coexist, leading to what is known as the Kerr-Taub-NUT spacetime~\cite{osti_4444917}.

Studies on physical implications of NUT charges can be found in the literature, especially in the context of astrophysical observations. 
The investigations based on geodesic analysis are: Lensing and atomic spectra~\cite{LyndenBell:1996xj}, gravitational lensing~\cite{NouriZonoz:1998va,Wei:2011nj}, microlensing~\cite{Rahvar:2002es,Rahvar:2003fh}, particle acceleration~\cite{Shen:2003pi,Liu:2010ja}, circular geodesics~\cite{Pradhan:2014zia}, and electromagnetic radiation from near horizon region in Kerr-Taub-NUT spacetime~\cite{Long:2018tij}. A complete analysis of geodesics, singularities, and causal structures of Taub-NUT spacetime has also been done recently~\cite{Kagramanova:2010bk}. Perfect fluids that can source NUT charge have been studied in ref.\cite{GarciaReyes:2004qn}. For a possibility of actual astrophysical objects having NUT charges, ref.\cite{Chakraborty:2017nfu} suggested the stellar-mass black hole-candidate X-ray binary, GRO J1655–40, as a possible NUT-charged object. Since horizons exist in Taub-NUT spacetime, black hole thermodynamics can be applied to it as well. This has been studied in recently papers such as refs. \cite{Bordo:2019rhu,Durka:2019ajz, Kubiznak:2019yiu,Bordo:2020kxm,Awad:2020dhy}, although the thermodynamics of their Euclidean cousins have been studied much earlier~\cite{Gibbons:1979xm}.

In the aforementioned studies the NUT parameter was predominantly assigned to the static background, and to best of authors' knowledge any detailed study on consequences of dynamical objects (in the sense that they can be accelerated or decelerated by external influences) having NUT charges has not yet appeared. This is where the recent proposals of understanding the NUT charges as "electric-magnetic duality transformations" of graviton coupling in the context of scattering amplitudes~\cite{Huang:2019cja,Emond:2020lwi} become interesting, as the proposal can be used to study dynamical properties of NUT-charged particles.

The works, refs.\cite{Huang:2019cja,Emond:2020lwi,Moynihan:2020gxj}, have appeared as attempts to reconcile electric-magnetic duality of gauge theory~\cite{Montonen:1977sn,Seiberg:1994rs,Seiberg:1994aj, Olive:1995sw, Harvey:1996ur, Olive:1997fg} with double copy relations of gauge theory and gravity~\cite{Kawai:1985xq,Bern:2008qj, Bern:2010ue,Bern:2010yg,Monteiro:2014cda}; this is the same motivation behind related works, refs.\cite{Luna:2015paa,Alawadhi:2019urr,Banerjee:2019saj,Bahjat-Abbas:2020cyb}, but the latter is restricted to the solutions of classical field equations\footnote{Consult refs. \cite{Ridgway:2015fdl,Carrillo-Gonzalez:2017iyj,Goldberger:2017vcg,Bahjat-Abbas:2017htu,Lee:2018gxc,Kim:2019jwm,delaCruz:2020bbn} and references therein for details on classical double copy.}. The works were based on the observation that a phase rotation to the kinematic factor $x$ appearing in on-shell three-point kinematics of equal mass-massless($m_1=m_2=m, m_3=0$) scattering, defined as the proportionality constant~\cite{Arkani-Hamed:2017jhn},
\bl
m x \l^\a &= \bar{\l}_{\dot\a} p_1^{\dot\a \a} \,,
\el
where $\l^\a$ and $\bar{\l}_{\dot\a}$ are spinor-helicity variables of the massless leg, generates electric-magnetic duality transformations on Maxwell's theory.

From the perspectives of an on-shell photon, electric and magnetic charges are distinguished by how they couple to distinct helicity modes of the photon; electric charges couple with the same sign, while magnetic charges couple with the opposite sign~\cite{Weinberg:1965rz}. Reconciling the sign choices with CPT symmetry forces magnetic couplings to have imaginary coupling constants~\cite{Weinberg:1965rz}, and the phase rotation to the kinematic factor $x$ introduced in ref.\cite{Huang:2019cja} can be understood as a phase rotation to the coupling constant that generates magnetic couplings from electric couplings. Pushing the analogy to gravitons, an ordinary mass would couple to both helicity modes of the gravitons with the same sign while a "magnetic mass" would couple with opposite signs. However, no known current could be found in Weinberg's analysis that would "magnetically" couple with the graviton and be consistent with CPT symmetry~\cite{Weinberg:1965rz}, and for this reason the possibility had been ruled out. Nevertheless, if such a coupling exists, it would correspond to the NUT parameter of (Kerr-)Taub-NUT spacetime~\cite{Huang:2019cja,Emond:2020lwi}, at least to leading Newton's constant order ($G^1$) in perturbation theory. The goal of present work is to examine this relation at next-to-leading order ($G^2$) in perturbation theory.

The manuscript is structured as follows. The motion of a test particle on Taub-NUT spacetime determined by general relativity will be reviewed in section \ref{sec:GR} to the second order in Newton's constant, referred to as the second post-Minkowskian (2PM) order in the literature. The \emph{deflection angle}\footnote{The terminology \emph{scattering angle} is also used in the literature, but the name will be reserved for a related observable. This observable is free of gauge ambiguities (general covariance) of general relativity.} will be the main observable that will be compared with quantum field theory computations. The scattering amplitude computation of the equivalent perturbation order will be performed in section \ref{sec:amplitude}. In section \ref{sec:boxes} we find that the widely-used approximation tool, the \emph{eikonal approximation}, is not valid at one-loop order when electric-magnetic duality transformation has been performed. Nonetheless, the imaginary part of the impulse computations cancel and a sensible deflection angle can be obtained. Based on a na\"ive expectation for the Compton amplitude, we obtain the deflection angle in section \ref{sec:triangles} and find a disagreement with the predictions of general relativity. Possible resolutions for closing the gap are considered in sections \ref{sec:ComptonFix} and \ref{sec:excitations}, which turn out to be insufficient for restoring coherence. As a by-product of our analysis, we obtain a relation between tidal response parameters and resonance excitation parameters, the equations \eqc{eq:transmute2tidal} and \eqc{eq:transmute2tidal2}. We conclude our study in section \ref{sec:conc}.

\section{The scattering angle from GR geodesic motion} \label{sec:GR}

The Taub-NUT metric is given as
\beq
ds^{2}=-&f(r;M,N)(dt+2N\cos{\th}d\vph)^2+\frac{dr^2}{f(r;M,N)}+(r^2+N^2)d\W_{S^2}^{2},\label{eq:metric}
\\
&f(r;M,N)\equiv\frac{r^2-2Mr-N^2}{r^2+N^2},
\qquad r_{h}=M+\sqrt{M^2+N^2},
\eeq
where $M$, $N$, and $r_h$ are ADM mass, NUT parameter(charge), and the location of the outer horizon, respectively. The Cauchy horizon lies outside the domain $r\in[0,\infty)$ and therefore can be ignored for our purposes. Regularity of spacetime requires periodicity of $2\p/N$ for the time-like $t$ coordinates, but the condition can be loosened when dual supertranslation symmetry is promoted to a gauge symmetry~\cite{Kol:2020ucd}. The Euclidean version of the solutions known as \emph{Euclidean Taub-NUT} and \emph{Taub-Bolt} solutions, which appear in the context of gravitational instantons~\cite{Gibbons:1979xm}, are obtained from the above metric by a Wick rotation, $t\to-i\t$ and $N\to -iN$, and their differences arise from allowed range for the radial coordinate $r$.

\subsection{Killing fields, integrals of motion, and geodesics}
Killing vector fields are the generators of the Taub-NUT spacetime isometry group, which in the coordinate patch of (\ref{eq:metric}) take the following form \cite{Visinescu:1999zs}.
\beq
\begin{split}
K^{(0)}=\pt_{t},\\
K^{(3)}=\pt_{\vf},   
\end{split}
&&
\begin{split}
\left(\begin{array}{c}
K^{(1)}
\\
K^{(2)}
\end{array}\right)=\left(\begin{array}{c}
\cos{\vf}
\\
\sin{\vf}
\end{array}\right)\left[2N\fr{1}{\sin{\th}}\pt_{t}-\cot{\th}\pt_{\vf}\right]+\pt_{\vf}\left(\begin{array}{c}
\cos{\vf}
\\
\sin{\vf}
\end{array}\right)\pt_{\th}\,.
\end{split} \label{eq:Killing1}
\eeq
Taking the inner product with the tangent vector $\partial_\l$ gives the integrals of motion for the geodesics.
\beq
\CE&=-g(\pt_{\l},K^{(0)})=f\fr{dt}{d\l}+2Nf\cos{\th}\fr{d\vf}{d\l},\nn
\\
J_{\vf}&=g(\pt_\l,K^{(3)})=-2N\CE\cos{\th}+(r^2+N^2)\sin^{2}{\th}\dot{\vf},
\\
C&=\left(\begin{array}{c}
g(\pt_\l,K^{(1)})
\\
g(\pt_\l,K^{(2)})
\end{array}\right)=\left[-\fr{2N\CE+J_{\vf}\cos{\th}}{\sin{\th}}+(r^2+N^2)\fr{d\th}{d\l}\pt_{\vf}\right]\left(\begin{array}{c}
\cos{\vf}
\\
\sin{\vf}
\end{array}\right).\nn
\eeq
The relations can be inverted to find the tangent vector,
\beq
\dot{\vf}=\fr{2N\CE\cos{\th}+J_{\vf}}{(r^2+N^2)\sin^{2}{\th}},&&
\dot{t}=\fr{\CE}{f}-2N\cos{\th}\dot{\vf},&&
\dot{\th}^{2}=\fr{\CC^2 \sin^{2}{\th}-(2N\CE+J_{\vf}\cos{\th})^{2}}{(r^2+N^2)^{2}\sin^{2}{\th}}, 
\eeq
where $\CC^{2}=C^{2}_{1}+C^{2}_{2}$. The geodesic equation for a particle of unit mass then reduces to
\beq
\CE\fr{dt}{d\l}-J_{\vf}\fr{d\vf}{d\l}-(r^2+N^2)(\fr{d\th}{d\l})^{2}-\fr{1}{f}(\fr{dr}{d\l})^{2}=1,
\eeq
which can be further reduced to the one-dimensional radial equation,
\beq
\left(\fr{dr}{d\l}\right)^{2}=\CE^2-V_{\textrm{eff}}, &&V_{\textrm{eff}}=\frac{r^2-2Mr-N^2}{r^2+N^2}\left(\fr{\CC^2+J_{\vf}^{2}-4N^2 \CE^2}{(r^2+N^2)}+1\right). \label{eq:radial1}
\eeq
The parameters can be chosen without loss of generality such that the geodesic lies on an \emph{orbital cone}\footnote{The conditions $\CC>0$ and $\CC=0$ correspond to linear and spiral scattering, respectively~\cite{Kagramanova:2010bk}.} of $\th=\th_0$ determined as
\beq
\cos{\th_0}=-\frac{2N\CE}{J_{\vf}}. \label{eq:coneangle1}
\eeq
Unfortunately this coordinate system is inconvenient for the Hamilton-Jacobi analysis as the method of separation of variables cannot be applied. Doing a gauge transformation allows application of the method for geodesics lying on the orbital cone $\th=\th_0$.

\subsection{Gauge transformation and action-angle variables}
The Hamilton-Jacobi analysis can be made more tractable by the gauge transformation, 
\beq
t\rightarrow t_{\OC}=t+2N\cos{\th_0}\vf_{\OC},&&\vf\rightarrow\vf_{\OC}=\vf.
\eeq
The metric \eqc{eq:metric} is then transformed to
\beq
ds^{2}=-f(r;M,N)(dt_{\OC}+2N(\cos{\th}-\cos{\th_{0}})d\vph)^2+\frac{dr^2}{f(r;M,N)}+(r^2+N^2)d\W_{S^2}^{2} \,,
\eeq
while the Killing fields \eqc{eq:Killing1} become
\beq
\begin{split}
K^{(0)}_{\OC}&=\pt_{t_{\OC}},
\\
K^{(3)}_{\OC}&=\pt_{\vf_{\OC}},
\end{split}
&&
\begin{split}
K_{\OC}^{(1)}=2N\cos{\vf}\fr{(1-\cos{\th}\cos{\th_0})}{\sin{\th}}\pt_{t}-\cos{\vf}\cot{\th}\pt_{\vf}-\sin{\vf}\pt_{\th},
\\
K_{\OC}^{(2)}=2N\sin{\vf}\fr{(1-\cos{\th}\cos{\th_0})}{\sin{\th}}\pt_{t}-\sin{\vf}\cot{\th}\pt_{\vf}+\cos{\vf}\pt_{\th} \,.  
\end{split} \label{eq:Killing2}
\eeq
The old Killing fields \eqc{eq:Killing1} and the new Killing fields \eqc{eq:Killing2} are related by the relations,
\beq
\begin{split}
K^{(0)}&=K_{\OC}^{(0)},
\\
K^{(3)}&=K^{(3)}_{\OC}+2N\cos{\th_0}K^{(0)}_{\OC},   
\end{split}
&&
\begin{split}
K^{(1)}&=K_{\OC}^{(1)},
\\
K^{(2)}&=K_{\OC}^{(2)}.
\end{split}
\eeq
Integrals of motion from $K_{\OC}$ are given as
\beq
\begin{split}
\CE&=f\dot{t}_{\OC}+2Nf(\cos{\th}-\cos{\th_0})\dot{\vf},
\\
J&=-2N(\cos{\th}-\cos{\th_0})\CE+(r^2+N^2)\sin^{2}{\th}\dot{\vf},
\\
C&=\left[-\fr{2NE(1-\cos{\th}\cos{\th_{0}})+J_{\OC}\cos{\th}}{\sin{\th}}+(r^2+N^2)\fr{d\th}{d\l}\pt_{\vf}\right]\left(\begin{array}{c}
\cos{\vf}
\\
\sin{\vf}
\end{array}\right).
\end{split} \label{eq:IntMotNew}
\eeq
The tangent vector components\footnote{The condition $\CC=0$ needs to be imposed for non-straight-line scattering motion.} transform as
\beq
\begin{split}
\dot{\vf}&=\fr{2N\CE(\cos{\th}-\cos{\th_0})+J}{(r^2+N^2)\sin^{2}{\th}},
\\
\dot{t}_{\OC}&=\fr{\CE}{f}-2Nf(\cos{\th}-\cos{\th_0})\dot{\vf},
\end{split}
&&
\dot{\th}^{2}=-\left(\fr{2N\CE(1-\cos{\th}\cos{\th_{0}})+J\cos{\th}}{(r^2+N^2)\sin{\th}}\right)^{2} ,
\eeq
and the geodesic equation reduces to
\beq
f^{-1}\CE^{2}-\fr{J^{2}}{(r^2+N^2)\sin^{2}{\th_0}}-f^{-1}\dot{r}^2=1.
\eeq
We remind the reader that the geodesic lies on the orbital cone, $\th=\th_0$, and the metric can be effectively considered as three-dimensional;
\beq
ds^{2}&=-fdt_{\OC}^2+\frac{dr^2}{f}+(r^2+N^2)\sin^{2}{\th_0}d\vf^2.
\eeq
The radial equation \eqc{eq:radial1} becomes
\beq
\dot{r}^2=\CE^{2}-V_{\eff},
&&
V_{\eff}=f\left(\fr{J^{2}}{(r^2+N^2)\sin^{2}{\th_0}}+1\right).
\eeq
The opening angle of the orbital cone in terms of the parameters of the geodesic, \eqc{eq:coneangle1}, needs to be expressed using new integrals of motion \eqc{eq:IntMotNew}. We solve the pair of equations,
\beq
\cos{\th_0}=-\frac{2N\CE}{J_{\vf}},&&J_{\vf}=J-2N\CE \cos{\th_0},
\eeq
for $J_{\vf}$ and obtain
\beq
J_{\vf}=\frac{J+\sqrt{J+16N^2 \CE^2}}{2} \,,
\eeq
with the sign choices $J_{\vf}>0$, $J>0$, $N>0$, and $\cos{\th_0}<0$. The opening angle $\th_0$ in terms of $\CE$, $J$ and $N$ is then given as
\beq
\cos{\th_0}=-\frac{4 N \CE}{J+\sqrt{J^2+16 N^2 \CE^2}},&&\sin^{2}{\th_0}=\frac{2 J}{\sqrt{J^{2}+16 N^2 \CE^{2}}+J}.
\eeq

\subsection{Hamilton-Jacobi analysis on the orbital cone}
The action functional in terms of action-angle variables on the orbital cone can be written using the effective three-dimensional metric.
\beq
\CS=-\CE t_{\OC}+J\vf+\int P(r')dr'=-\m \int d\l,
\eeq
where $\m$ is the mass of the probe particle moving along the geodesic. The relativistic Hamilton-Jacobi equation is
\beq
g^{\m\n} \pt_\m \mathcal{S} \pt_\n \CS =-\mu^{2},
\eeq
which coincides with the geodesic equation on the orbital cone.
\beq
-f^{-1}\CE^{2}+\frac{J^{2}}{(r^{2}+N^{2})\sin^{2}{\th_0}}+fP(r)^{2}=-\m^{2}.
\eeq
Solving for the radial momentum $P(r)$ gives
\beq
P(r)=\pm\left[f(r;M,N)^{-2}\CE^{2}-\frac{\m^{2}}{f(r;M,N)}-\frac{J^{2}}{f(r;M,N)(r^{2}+N^{2})\sin^{2}{\th_0}}\right]^{\half}.
\eeq
Note that $f^2 P^2=\CE^{2}-V_{\eff}$. 

\subsection{Scattering at second post-Minkowskian order}
Since we are interested in effects up to second post-Minkowskian order (2PM) or to second order in Newton's constant $G$, the expansion parameter $G$ will be restored to $GM$ and $GN$ in this section. Standard Hamilton-Jacobi analysis can be applied to compute the angle variables. The change in the azimuthal angle during scattering, $\D\vf$, is
\beq
\D\vf=-2\int_{r_{\textrm{min}}}^{\infty}\frac{\pt P(r)}{\pt J}dr=-2\int_{u_{\textrm{min}}}^{\infty}\frac{\pt P(r(u))}{\pt J}du=2(\vf(\infty)-\vf(b)),\label{eq:Azi}
\eeq
where the $r_{\min}$ is the largest positive root of the equation $\CE^2-V_{\eff}=0$. Expanding $r_{\min}$ to $G^2$ order,
\beq
r_{\min}=\frac{J}{\sqrt{\mathcal{E}^2-\mu ^2}}+\frac{G M \mathcal{E}^2}{\mu ^2-\mathcal{E}^2}+\frac{G^2 \left(M^2 \left(4 \mu ^2 \mathcal{E}^2-3 \mathcal{E}^4\right)+N^2 \left(\mathcal{E}^4-\mu ^4\right)\right)}{2 J \left(\mathcal{E}^2-\mu ^2\right)^{3/2}}+\CO(G^3).
\eeq
When evaluating the integral \eqc{eq:Azi} to $G^2$ order, simply expanding the integrand and lower/upper limits to $G^2$ order and effecting the integral mechanically will not give correct results. This is because the perturbations of integral limits can ruin the structure of the integral. Reparametrising the integration variable so that the integral limits are not perturbed will make the perturbation integral well-behaved.

The change in azimuthal angle during scattering is then given as 
\beq
\D\vf&=\pi+\frac{2 M \left(2\CE^2-\mu ^2 \right)}{\sqrt{\CE^2-\mu ^2}}\left(\frac{ G }{J}\right)\nn
\\
&\qquad-\frac{\pi \left(3 \mu ^2 \left(M^2+N^2\right)+\CE^2 \left(N^2-15 M^2\right)\right)}{4}\left(\frac{ G }{J}\right)^{2}+O\left(G^3\right).
\eeq
Due to the presence of NUT parameter $N$, the geodesic motion lies on the orbital cone of $\th=\th_0$ and $\D\vf$ should not be considered as the scattering angle. Instead, the \emph{scattering angle} $\F$ and the \emph{deflection angle} $\vth$ are given as
\beq
\cos{\F}=\cos^{2}{\th_{0}}+\sin^{2}{\th_{0}}\cos{\D\vf},&&\vth=\F-\p.\label{eq:AziToSca}
\eeq
Solving the above equation to $G^2$ order yields the 2PM scattering angle
\beq
\F&=\pi+\vth=\pi +2\sqrt{4\CM_{\TN}^{2}\CE^{2}+\frac{M^{2}\m^{4}}{\CE^2-\mu ^2}}\left(\frac{ G }{J}\right)\nn
\\
&\qquad+\pi\frac{ M  \left(2 \CE^2-\mu ^2\right)\left(16 M^2 \CE^2-\CM_{\TN}^{2} \left(3 \mu ^2+\CE^2\right)\right)}{4\sqrt{\mu ^4 M^2+4 \CM_{\TN}^{2}\CE^{2}(\CE-\mu ) (\CE+\mu)}}\left(\frac{ G }{J}\right)^{2}+\CO(G^3).\label{eq:defangle}
\eeq
where $\CM_{\TN}^{2}=M^{2}+N^{2}$. In the Schwarzschild limit, $N \to 0$, the orbital cone opening angle $\th_0$ becomes $\pi/2$ and geodesic motion becomes equatorial. In this limit $\D\vf$ coincides with $\F$.
\beq
\D\vf_{{\rm Sch}}&=\F_{{\rm Sch}}=\pi+\frac{2M \left(2 \CE^2-\mu ^2\right)}{ \sqrt{\CE^2-\mu ^2}}\left(\frac{ G }{J}\right)+\frac{3 \pi M^2 \left(5 \CE^2-\mu ^2\right)}{4}\left(\frac{ G }{J}\right)^{2}+\CO\left(G^3\right).\label{eq:schangle}
\eeq

The deflection angle $\vth$ is vectorial in this set-up, therefore we define the \emph{deflection angle vector} $\vec{\vth}$ as $\hat{n}_{\rm out}-\hat{n}_{\rm in}$. This is the main observable that will be compared with scattering amplitude source-probe limit computations.
\beq
\vec{\vth}=\left(\begin{array}{c}
-\cos^{2}{\D\th_0}\left(\cos{\vth_{\vf}}-1\right)-2\sin^{2}{\D\th_0}
\\  
-\cos{\D\th_0}\sin{\vth_\vf}
\\
-2\sin{\D\th_0}\cos{\D\th_0}(1+\sin^{2}{\vth_{\vf}})
\end{array}\right),
\eeq
where $\D\th_0=\th_0-\p/2$, $\vth_{\vf}=\D\vf-\p$. The deflection angle $\vth$ is given as the size of the deflection angle vector $\vth = \abs{\vec{\vth}}$. In amplitude computations, the incoming test particle will be considered as directed along the $x$-axis, which is \emph{not} the incoming direction of the geodesic analysis. Therefore we rotate the reference frames to match the normalisations of amplitude computations.
\beq
\begin{split}
\vec{\vth}&=\hat{n}^{\rm amp}_{\rm out}-\hat{n}^{\rm amp}_{\rm in}=\CR_{\D\th_0} \vec{\vth}_{\rm GR},
\\
\vec{\vth}_{\rm GR}&=\hat{n}^{\rm GR}_{\rm out}-\hat{n}^{\rm GR}_{\rm in},
\end{split}
&&
\begin{split}
\hat{n}^{\rm amp}_{\rm in}&=\CR_{\D\th_0} \hat{n}^{\rm GR}_{\rm in}=-\hat{x},
\\
\hat{n}^{\rm amp}_{\rm out}&=\CR_{\D\th_0} \hat{n}^{\rm GR}_{\rm out}.
\end{split}
\eeq
The vectors and matrices can be explicitly evaluated.
\beq
\CR_{\D\th_0}=\left(\begin{array}{ccc}
  \cos{\D\th_0}   & 0& \sin{\D\th_0}\\
    0 & 0& 0\\
  -\sin{\D\th_0}   & 0&\cos{\D\th_0}
\end{array}\right),&&
 \hat{n}^{\rm GR}_{\rm in}=\left(\begin{array}{c}
  -\cos{\D\th_0} 
   \\
   0
   \\
\ph{-}\sin{\D\th_0} 
 \end{array}\right)
 ,&&
 \hat{n}^{\rm GR}_{\rm out}=\left(\begin{array}{c}
  -\cos{\D\th_0}\cos{\vth_\vf} 
   \\
-\cos{\D\th_0} \sin{\vth_\vf}
\\
-\sin{\D\th_0} 
 \end{array}\right).
\eeq
The 2PM deflection angle vector in the transformed frame is then given as
\beq
\vec{\vth}=\left(\begin{array}{c}
0
\\  
\frac{2 M \left(\mu ^2-2 \mathcal{E}^2\right)}{ \sqrt{\mathcal{E}^2-\mu ^2}}
\\
-4N \mathcal{E}
\end{array}\right)\left(\frac{ G}{J}\right)
+
\left(\begin{array}{c}
- 4\left(4\CM_{\TN}^{2}\mathcal{E}^2-8 M^2 \mathcal{E}^2-\frac{M^2 \mu ^4 }{\mathcal{E}^2-\mu ^2}\right)
\\  
\pi\frac{ \CM_{\TN}^{2}\left(\mathcal{E}^2 +3 \mu ^2\right)-16 M^2 \mathcal{E}^2 }{4}
\\
0
\end{array}\right)\left(\frac{ G}{J}\right)^{2}+\CO\left(G^3\right) \,.\label{eq:2pmAnglevec}
\eeq
The Schwarzschild limit can be recovered by the limit $N \to 0$. The first component of the vector is directed along the direction of the incoming test particle, and is irrelevant for the perturbation order we are interested in. Note that while electric ($M$) and magnetic ($N$) masses deflect the particle in orthogonal directions at 1PM order ($G^1$), the deflections are aligned at 2PM order($G^2$).

\section{The classical deflection angle from QFT amplitudes} \label{sec:amplitude}

\subsection{Momentum parametrisation and the holomorphic classical limit}
A generic one-loop amplitude can be expanded on a basis of integrals known as \emph{scalar integrals}. Unitarity methods~\cite{Bern:1994zx,Bern:1994cg} aim to compute the coefficients of the basis scalar integrals by examining analytic structures of the loop integral, often building up the integrand from lower perturbation order on-shell amplitudes. We resort to unitarity methods for the analysis of one-loop amplitude, as the analysis of refs.\cite{Huang:2019cja,Emond:2020lwi} was based on on-shell tree amplitudes whose continuation to off-shell kinematics\footnote{while retaining Lorentz invariance and locality} is unclear; the textbook method of constructing the loop integrand from Feynman rules requires off-shell data. 

Classical physics is dictated by long-distance physics, or physics at large impact parameters. Since impact parameter $b^\m$ is the Fourier dual of transferred momentum $q^\m$ on the impact parameter space, the expansion in large impact parameter can be traded for an expansion in small transferred momentum. Dimensional analysis in Planck's constant $\hbar$ shows that only the scalar box integrals and the scalar triangle integrals have the relevant dimensions to contribute to classical physics~\cite{Holstein:2004dn,Neill:2013wsa,Guevara:2017csg,Bern:2019crd}.

A computationally efficient means of probing the leading contributions is the \emph{holomorphic classical limit} (HCL)~\cite{Guevara:2017csg}, where transfer momentum $q^\m$ is complexified to a non-zero complex null vector. The subdominant $q^2$ contributions drop out in the computations as $q^\m$ is a null vector, but as $q^\m$ is non-zero certain information such as spin dependence can be reconstructed in this limit~\cite{Chung:2018kqs,Chung:2019duq,Chung:2019yfs,Chung:2020rrz,Guevara:2018wpp,Guevara:2019fsj,Moynihan:2019bor}. Cut momenta parametrisation will be given in this section for later reference; consult ref.\cite{Guevara:2017csg} for derivation of the cut momenta parametrisations.

The parametrisation for external momenta is given as
\bl
\bld
p_1 &= ( E_a, \vec{p} + \vec{q} / 2 ) = \sket{\hat\eta} \bra{\hat\l} + \sket{\hat\l}\bra{\hat\eta} \,,
\\ p_1' &= ( E_a, \vec{p} - \vec{q} / 2 ) = \b' \sket{\hat\eta} \bra{\hat\l} + \frac{1}{\b'} \sket{\hat\l}\bra{\hat\eta} + \sket{\hat\l} \bra{\hat\l} \,,
\\ p_2 &= ( E_b, - \vec{p} - \vec{q} / 2 ) = \sket{\eta} \bra{\l} + \sket{\l}\bra{\eta} \,,
\\ p_2' &= ( E_b, - \vec{p} + \vec{q} / 2 ) = \b \sket{\eta} \bra{\l} + \frac{1}{\b} \sket{\l}\bra{\eta} + \sket{\l} \bra{\l} \,,
\\ q &= p_1 - p_1' = (0,\vec{q}) \,,
\eld
\el
where spinor brackets are normalised by the conditions\footnote{$m_a$ and $m_b$ are inertial masses. The notation is summarised in Table \ref{tab:notations}.}, $\la \hat\l \hat\eta \ra = [\hat\l \hat\eta] = m_a$ and $\la \l \eta \ra = [\l \eta] = m_b$. The limit $\b \to 1$ corresponds to the HCL, and following definitions for the variables will be used.
\begin{table}
    \centering
\begin{tabular}{|c|c|c|c|}
\hline
Inertial Mass&$\CM^2=M^2+N^2$&$m_a\to \CM_a$&$m_b\to\CM_b$  \\
\hline
Electric Mass&$M$&$M_a$&$M_b$
\\
\hline
Magnetic Mass&$N$&$N_a$&$N_b$
\\
\hline
\end{tabular}
\caption{Gravitational charges}
\label{tab:notations}
\end{table}
\bl
\bld
u = \sbra{\l} p_1 \ket{\eta} \,,\, v = \sbra{\eta} p_1 \ket{\l} \,,\, \s = \frac{p_1 \cdot p_2}{m_a m_b} \,.
\eld
\el
The momentum $p_1$ can be expanded in terms of unhatted spinors using above variables\footnote{The expansion can be derived using eq.(6.10) of ref.\cite{Chung:2018kqs}.}.
\bl
\bld
p_1 &= \sket{\eta} \frac{u}{m_b^2} \bra{\l} + \sket{\l} \frac{v}{m_b^2} \bra{\eta} + \frac{\sket{\l} \b (uv - m_a^2 m_b^2) \bra{\l}}{A {m_b^2}} + \frac{\sket{\eta} A \bra{\eta}}{\b {m_b^2}} \,,
\\ A &= (\b-1)(\b v - u + (\b-1) m_b^2) \,.
\eld
\el
\bfig
\centering
\includegraphics[width=0.4\textwidth]{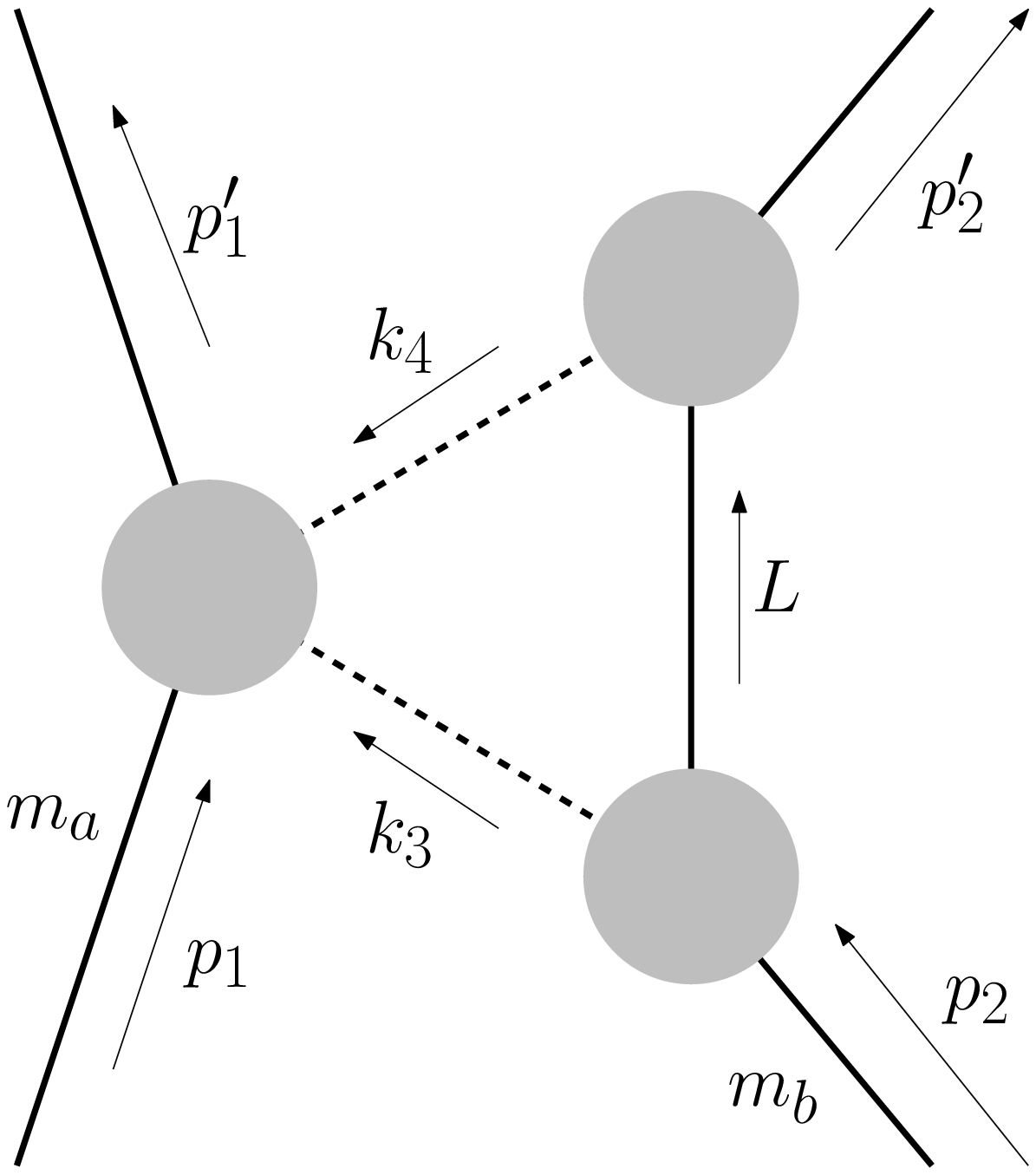} \caption{The $b$-topology triangle cut with cut conditions $k_3^2 = k_4^2 = L^2 - m_b^2 = 0$. The $a$-topology cut is defined as the mirror topology obtained by exchanging $m_a \leftrightarrow m_b$.} \label{fig:b-top_cut}
\efig
For the $b$-topology triangle cut, \fig{fig:b-top_cut}, the cut momenta are given as
\bl
\bld
\ket{k_3} = \frac{1}{\b + 1} \left( (\b^2 -1) \ket{\eta} - \frac{1 + \b y}{y} \ket{\l} \right) &\,,\,\sket{k_3} = \frac{1}{\b + 1} \left( (\b^2 -1) y \sket{\eta} + (1 + \b y) \sket{\l} \right) \,,
\\ \ket{k_4} = \frac{1}{\b + 1} \left( \frac{\b^2 - 1}{\b} \ket{\eta} + \frac{1-y}{y} \ket{\l} \right) &\,,\, \sket{k_4} = \frac{1}{\b + 1} \left( - \b(\b^2 - 1) y \sket{\eta} + (1 - \b^2 y) \sket{\l} \right) \,. \label{eq:cutmom}
\eld
\el
The $b$-topology contribution to the amplitude can be computed as
\bl
\bld
i \CA^{\bigtriangleup^{(b)}} &= - I_{\bigtriangleup^{(b)}} \oint_{\G} \frac{dy}{2 \pi i y} \left[\sum_{h_3,h_4} \CA_4 \times \CA_3 \times \CA_3\right] (y) \,,
\\ I_{\bigtriangleup^{(b)}} &= \frac{-i}{32 m_b \sqrt{-q^2}} + \cdots \,,
\eld \label{eq:TriCutSol}
\el
where integration contour $\G$ for $y$ encircles $y=\infty$, which can be traded for reading off $y^0$ coefficient of the expression's Laurent expansion at $y=\infty$~\cite{Forde:2007mi}. This scalar triangle integral contribution captures the $\abs{\vec{q}}^{-1} \sim r^{-2}$ dependence of the classical Hamiltonian in position space and $\abs{\vec{q}}^{-1} \sim b^{-1}$ dependence of the eikonal phase in impact parameter space.

\subsection{The impulse formula}
Define \emph{pre-eikonal phase} as
\bl
\bld
\chi [\CA] (b , p_1, p_2) &:= \int \hat{d}^4 q \hat{\delta} (2p_1 \cdot q + q^2) \hat{\delta} (2p_2 \cdot q - q^2) \Th (p_1^0 + q^0) \Th (p_2^0 - q^0)
\\ &\phantom{asdfasdfasdf} \times e^{-i b \cdot q} \CA (p_1, p_2 \to p_1+q,p_2-q) \,,
\eld
\el
where
\bl
\hat{d}^n p := \frac{d^n p}{(2\pi)^n} \,,\, \hat{\delta}(x) := 2 \pi \delta (x) \,,
\el
are the normalised measure and delta constraint in momentums space. While the above integral kernel is widely used to compute eikonal amplitudes, the terminology pre-eikonal phase has been adopted because diagrams that do not contribute to the eikonal phase will also be effected by this kernel. In the COM frame with small momentum mismatch $q^\m \sim w_1^\m \sim 0$, the delta constraints can be simplified as
\bl
\chi [\CA] (b , p_1, p_2) &\stackrel{q^2 \to 0}{\simeq} \frac{1}{4 \CE_{real} \abs{\vec{p}_{\COM}}} \int \hat{d}^2 q \, e^{i \vec{b} \cdot \vec{q}} \CA (p_1, p_2 \to p_1+q,p_2-q) \,, \label{eq:HCLPEP}
\el
where $\CE_{real} = \sqrt{s}$ is the COM frame total energy and $\vec{p}_{\COM}$ is the spatial momentum in the COM frame. The above integral is performed in the Euclidean 2-plane dual to the impact parameter space $\vec{b}$. Note that
\bl
\abs{\vec{p}_{\COM}} &= \frac{m_a m_b \sqrt{\s^2 - 1}}{\sqrt{s}} \,,
\el
therefore the normalisation factor for the deflection angle is given as~\cite{Guevara:2018wpp}
\bl
\frac{1}{4 \CE_{real} \abs{\vec{p}_{\COM}}^2} &= \frac{\CE_{real}}{(2 m_a m_b \sqrt{\s^2 - 1})^2} \,.
\el
The impulse formula eq.(4.31) and eq.(4.32) of ref.~\cite{Kosower:2018adc} can be approximated to
\bl
\Delta p_1^\m &\simeq - \frac{\partial \chi [\CA] (b, p1, p2)}{\partial b_\m} + i \chi [\CA^\ast] (-b, p1, p2) \frac{\partial \chi [\CA] (b, p1, p2)}{\partial b_\m} \,,
\el
when production of extra particles is neglected, i.e. $X = \varnothing$, and the full expression is assumed to be dominated by small momentum mismatch $q^\m \sim w_1^\m \sim 0$. This approximation is valid up to $G^2$ order in the classical limit, and the impulse up to this order can be written as;
\bl
\Delta p_1^\m &= - \frac{\partial \chi [\CA^{(0)+(1)}] (b)}{\partial b_\m} + i \chi [(\CA^{(0)})^\ast] (-b) \frac{\partial \chi [\CA^{(0)}] (b)}{\partial b_\m} + \CO (G^3) \,,
\el
where momenta of hard particles have been suppressed and the superscript on $\CA$ denotes loop-order of the amplitude; e.g. $\CA^{(n)}$ is the $n$-loop amplitude. Scaling the impulse by $\abs{\vec{p}_{\text{com}}}$ gives the deflection angle, which is reliable up to $G^2$ order\footnote{The reliability relies on two facts; (1) extra particle production can only contribute at $G^3$ or higher orders, and (2) the small angle approximation $2 \sin (\th/2) \simeq \th$ is exact up to $\th^2$ order, where leading scaling is $\th \propto G$.}.
\bl
\bld
\vec{\vth} &= \frac{\Delta p_1^\m}{\abs{\vec{p}_{
\text{com}}}} + \CO (G^3)
\\ &= \frac{\CE_{real}}{m_a m_b \sqrt{\s^2-1}} \left[ \frac{\partial \chi[\CA^{(0)+(1)}](b)}{\partial \vec{b}} - i \chi [(\CA^{(0)})^\ast] (-b) \frac{\partial \chi [\CA^{(0)}] (b)}{\partial \vec{b}} \right] + \CO (G^3) \,.
\eld
\el

The master integral for \eqc{eq:HCLPEP} is
\bl
\bld
I_{n,m}(\vec{b}) := \int \hat{\delta}^{n} q \abs{q}^m e^{i \vec{b} \cdot \vec{q}} &= \frac{2^{2-n} }{\pi^{\frac{n-3}{2}} \G[\frac{n-1}{2}] \abs{b}^{n+m}} \int_0^{\infty} \frac{dx}{2\pi} x^{n+m-1} J_0 (x)
\\ &= \frac{2^{m} \G[\frac{m+n}{2}]}{\pi^{\frac{n-1}{2}} \G[\frac{2-m-n}{2}] \G[\frac{n-1}{2}] \abs{b}^{m+n}} \,,
\eld \label{eq:MS1}
\el
where $n=2-2\e$ is the dimensional regulator for UV/IR divergences. The integral is convergent for $0<m+n<3/2$. Differentiating by $m$,
\bl
\bld
\frac{\partial I_{n,m}(\vec{b})}{\partial m} &= \int \hat{\delta}^{n} q \abs{q}^{m} \log \abs{q} e^{i \vec{b} \cdot \vec{q}} 
\\ &= \frac{I_{n,m}(\vec{b})}{2} \left[ -2 \log \abs{b} +\psi^{(0)}\left(\frac{2-m-n}{2}\right)+\psi^{(0)}\left(\frac{m+n}{2}\right)+\log (4)\right]\,.
\eld \label{eq:MS2}
\el
Following results will be used in the computations.
\bl
\left(\frac{4\m^{2}}{\p e^\g}\right)^{\e} I_{2-2\e,-2} &= -\frac{1}{4 \pi  \epsilon }-\frac{\log \abs{b \mu}}{2 \pi }+O\left(\epsilon ^1\right),
\\ I_{2,-1} &= \frac{1}{2 \pi \abs{b}} \,,
\\ \left(\frac{4\m^{2}}{\p e^\g}\right)^{\e} \left.\frac{\partial I_{2-2\e,m}}{\partial m} \right|_{m=-2} &= -\frac{1}{8 \pi  \epsilon ^2}+\frac{\gamma -\log (2 \mu )}{4 \pi  \epsilon } \nn
\\ &\phantom{=} +\frac{16(\gamma -\log (2)) \log \abs{b \mu}+8 \log ^2\abs{b}-8 \log ^2(\mu )+\pi ^2}{32 \pi }+O\left(\epsilon ^1\right) .
\el

\subsection{The impulse at tree order}

This analysis has already been presented in ref.\cite{Huang:2019cja}, which has been included for completeness. The on-shell three-point amplitudes are given as~\cite{Arkani-Hamed:2017jhn}
\bl
\CA_{\f\bar{\f}h}^{+} = \frac{\k m^2 (e^{i\vf}x)^2}{2} \,,\, \CA_{\f\bar{\f}h}^{-} = \frac{\k m^2}{2 (e^{i\vf}x)^2} \,, \label{eq:3pt}
\el
where $\k = \sqrt{32 \pi G}$, $\CA^{+}$($\CA^{-}$) is the coupling to positive(negative) helicity graviton, and $x$ is the kinematic factor carrying little group weights of the graviton.
\bl
x &= \frac{\sbra{3}p_1 \ket{\xi}}{m \la 3 \xi \ra} = \sqrt{2} p_1 \cdot \ve^{(+)}(p_3) \,, \label{eq:xdef}
\el
where $\ve^{(+)}_\m (p_3)$ is the positive helicity polarisation vector for a massless spin-1 particle with momentum $p_3$. The phase rotation $\vf$ generates magnetic coupling~\cite{Huang:2019cja}. The $m_a m_b \to m_a m_b$ tree amplitude can be obtained by gluing the on-shell three-point amplitudes, and when particle $a$ is subject to phase rotation $\vf$ the amplitude is given as;
\bl
\bld
\CA^{(0)} &\stackrel{q^2 \to 0}{\simeq} \frac{8 \pi G m_a^2 m_b^2}{q^2} \left[ \left(\frac{x_1}{x_2}\right)^2 e^{2 i \vf} + \left(\frac{x_2}{x_1} \right)^2 e^{-2i \vf} \right]
\\ &= \frac{16 \pi G m_a^2 m_b^2}{q^2} \left[ (2\s^2 - 1) \cos (2\vf) + 2 i \s \sqrt{\s^2 - 1} \sin (2\vf) \right] \,,
\\ \chi[\CA^{(0)}] &\simeq \frac{2 G m_a m_b}{\sqrt{\s^2-1}} \left[ (2\s^2 - 1) \cos (2\vf) + 2 i \s \sqrt{\s^2 - 1} \sin (2\vf) \right] \log\abs{b} \,.
\eld
\el
The sign difference comes from $q^2 = -\vec{q}^2 \leftrightarrow \frac{\log\abs{b\m}}{2\pi}$. The deflection angle from (pre-)eikonal phase $\chi$ is given as;
\bl
\bld
\vec{\vth}_1 &= \frac{\CE_{real}}{m_a m_b \sqrt{\s^2-1}} \frac{\partial \chi[\CA^{(0)}](b)}{\partial \vec{b}}
\\ &= \frac{\CE_{real}}{(2 \sqrt{\s^2-1})^2} \frac{8G}{\abs{b}} \left[ - (2\s^2 - 1) \cos(2\vf) \hat{y} + 2 \s \sqrt{\s^2 - 1} \sin(2\vf) \hat{z} \right] \,,
\eld
\el
where imaginary part of the impulse has been re-interpreted as $i\sqrt{\s^2-1} q^\m = \pm \e^\m(u_1,u_3,q)$, with sign choice fixed by matching to the Newtonian limit~\cite{Huang:2019cja}. This result can be matched to general relativity predictions \eqc{eq:2pmAnglevec} using the parameter mappings of table \ref{tab:mapping}, yielding
\bl
\vec{\vth}^{\rm Amp}_{1} &= -\left[ \frac{2M(2\CE^2-\m^2)}{(\sqrt{\CE^2-\m^2}) } \hat{y} + 4\CE N \hat{z} \right] \left(\frac{G}{J}\right) \,,
\el
consistent with \eqc{eq:2pmAnglevec} to this order.

\begin{table}
\centering
\begin{tabular}{|c|c|c|c|c|c|}
\hline
Amplitude variables&$\CE_{real}$ & $\s$&$b\sqrt{\s^2-1}$ &$\cos{(2\vf)}$&$\sin{(2\vf)}$
\\
\hline\hline
Geodesic variables&$\CM_{\TN}$&$\CE/\m$&$J/\m$&$M/\CM_{\TN}$&$N/\CM_{\TN}$\\
\hline
\end{tabular}
\caption{Parameter mappings between amplitude variables and geodesic variables. The mapping is based on the relations $\CE_{real} = \CM_{\TN} + \CE \simeq \CM_{\TN}$, $p_1 \cdot p_2 \simeq \CM_{\TN} \CE$, and $J = b \abs{\vec{p}_{\COM}}$.}
\label{tab:mapping}
\end{table}

\subsection{The one-loop amplitude : Boxes and eikonal (non-)exponentiation} \label{sec:boxes}
The box topology contribution to the one-loop amplitude in the HCL is;
\bl
i \CA^{\square} &\stackrel{q^2 \to 0}{\simeq} 64 \pi^2 G^2 m_a^4 m_b^4 (16\s^4 - 16\s^2 + 2  + e^{4i\vf} + e^{-4i\vf}) I_\square \,, \label{eq:boxcoeff}
\\ i \CA^{\triangleright \hskip -1pt \triangleleft} &\stackrel{q^2 \to 0}{\simeq} 64 \pi^2 G^2 m_a^4 m_b^4 (16\s^4 - 16\s^2 + 2  + e^{4i\vf} + e^{-4i\vf}) I_{\triangleright \hskip -1pt \triangleleft} \,, \label{eq:cboxcoeff}
\\ I_\square &= \frac{-i}{8 \pi^2} \frac{\log(-q^2)}{q^2} \frac{1}{2 m_a m_b \sqrt{\s^2 - 1}} \left[ \log \abs{\frac{\s-1-\sqrt{\s^2-1}}{\s-1+\sqrt{\s^2-1}}} + i \pi \right] \,,
\\ I_{\triangleright \hskip -1pt \triangleleft} &\stackrel{q^2 \to 0}{\simeq} \frac{-i}{8 \pi^2} \frac{\log(-q^2)}{q^2} \frac{1}{2 m_a m_b \sqrt{\s^2 - 1}} \left[ \log \abs{\frac{\s+1+\sqrt{\s^2-1}}{\s+1-\sqrt{\s^2-1}}} \right] \,,
\el
where $\s = \frac{p_1 \cdot p_2}{m_a m_b} > 1$ has been applied to the imaginary part of the scalar integrals. The scalar integral coefficients can be obtained by imposing the fourth cut condition on the triple-cut solutions \eqc{eq:cutmom}. Adding up, the logarithmic terms in square brackets cancel and only the imaginary part of the scalar integrals survive.
\bl
\bld
i \CA^{\square + \triangleright \hskip -1pt \triangleleft} &\stackrel{q^2 \to 0}{\simeq} \frac{4 \pi G^2 m_a^3 m_b^3 (16\s^4 - 16\s^2 + 2 + e^{4i\vf} + e^{-4i\vf})}{\sqrt{\s^2-1}} \frac{\log(-q^2)}{q^2} \,.
\eld
\el
The pre-eikonal phase is computed as;
\bl
\chi[\CA^{\square + \triangleright \hskip -1pt \triangleleft}] &\stackrel{q^2 \to 0}{\simeq} i \frac{2G^2 m_a^2 m_b^2 (16\s^4 - 16\s^2 + 2 + e^{4i\vf} + e^{-4i\vf})}{4 (\s^2-1)} \log^2 \abs{b} + \cdots \,,
\el
where only the leading $b$ dependence has been kept. Setting the phase rotation zero $\vf=0$,
\bl
\chi[\CA^{\square + \triangleright \hskip -1pt \triangleleft}] &\stackrel{q^2 \to 0}{\simeq} \frac{i}{2} \left[ \frac{2G m_a m_b (2\s^2 - 1)}{\sqrt{\s^2-1}} \log \abs{b} \right]^2 + \cdots \,,
\\ \chi[\CA^{(0)}] &\simeq \frac{2 G m_a m_b (2\s^2 - 1)}{\sqrt{\s^2-1}} \log \abs{b} \,,
\el
which is consistent with exponentiation of the eikonal phase; $\chi[\CA^{\square + \triangleright \hskip -1pt \triangleleft}] = \frac{i}{2} \chi[\CA^{(0)}]^2$. This relation obviously breaks down when non-zero phase rotation $\vf \neq 0$ is introduced, rendering eikonal approximation for dyonic amplitudes questionable. However, \emph{their impulse contributions cancel each other},
\bl
\frac{\partial \chi[\CA^{\square + \triangleright \hskip -1pt \triangleleft}](b)}{\partial \vec{b}} - i \chi [(\CA^{(0)})^\ast] (-b) \frac{\partial \chi [\CA^{(0)}] (b)}{\partial \vec{b}} &=0 \,,
\el
since
\bl
\bld
(16\s^4 - 16\s^2 + 2 + e^{4i\vf} + e^{-4i\vf}) - 4\abs{(2\s^2 - 1) \cos (2\vf) + 2 i \s \sqrt{\s^2 - 1} \sin (2\vf) }^2 &=0 \,. \nn
\eld
\el
Therefore,
\bl
\bld
\vec{\vth}_2 &= \frac{\CE_{real}}{m_a m_b \sqrt{\s^2-1}} \left[ \frac{\partial \chi[\CA^{\bigtriangleup + \bigtriangledown+\square + \triangleright \hskip -1pt \triangleleft}](b)}{\partial \vec{b}} - i \chi [(\CA^{(0)})^\ast] (-b) \frac{\partial \chi [\CA^{(0)}] (b)}{\partial \vec{b}} \right]
\\ &= \frac{\CE_{real}}{m_a m_b \sqrt{\s^2-1}} \left[ \frac{\partial \chi[\CA^{\bigtriangleup + \bigtriangledown}](b)}{\partial \vec{b}} \right] \,.
\eld
\el
Thus, only the triangle contributions contribute to the classical deflection angle.

\subsection{The one-loop amplitude : Triangles} \label{sec:triangles}
\subsubsection{Analysis from the na\"ive Compton amplitude}
The gravitational Compton amplitude is needed when evaluating the triangle contribution for the loop amplitude, as can be seen on the left side of \fig{fig:b-top_cut}. The gravitational Compton amplitudes are given as~\cite{Arkani-Hamed:2017jhn,Caron-Huot:2018ape,Chung:2018kqs,Johansson:2019dnu,Chung:2019duq}
\bl
\bld
\CA^{++}_{\f h h \bar{\f}} &= \frac{(\k/2)^2 m^4 [23]^4}{(s-m^2)t(u-m^2)} \,,
\\ \CA^{+-}_{\f h h \bar{\f}} &= \frac{(\k/2)^2 [2|p_1|3\ra^4}{(s-m^2)t(u-m^2)} \,,
\eld
\el
where Mandelstam invariants are defined as\footnote{All-outgoing convention is used.} $s=(p_1+p_2)^2$, $t=(p_1+p_4)^2$, and $u=(p_1+p_3)^2$. As in \eqc{eq:3pt}, NUT charges can be generated by phase rotation $e^{2i\vf}$($e^{-2i\vf}$) for each positive(negative) helicity graviton~\cite{Huang:2019cja}, which results in the expressions,
\bl
\bld
\CA^{++} &\to e^{4i\vf} \CA^{++} = e^{4i\vf} \frac{(\k/2)^2 m^4 [23]^4}{(s-m^2)t(u-m^2)} \,,
\\ \CA^{+-} &\to \CA^{+-} = \frac{(\k/2)^2 [2|p_1|3\ra^4}{(s-m^2)t(u-m^2)} \,. \label{eq:ComptonNaive}
\eld
\el
As noted in ref.\cite{Caron-Huot:2018ape}, these two amplitudes do \emph{not} satisfy the $t$-channel factorisation conditions\footnote{The $t$-channel pole does not exist for the electromagnetic Compton amplitude, therefore the analogous rotation generates a consistent dyonic amplitude for electromagnetism.}. This is the reason they are referred to as \emph{na\"ive Compton amplitudes}. Nevertheless, the impulse and the deflection angle will be analysed using the na\"ive amplitudes \eqc{eq:ComptonNaive} and the corrections due to the difference \eqc{eq:ComptonDiff} will be analysed in the following section.

Combining \eqc{eq:cutmom}, \eqc{eq:3pt}, and \eqc{eq:ComptonNaive} into \eqc{eq:TriCutSol}, it can be shown that
\bl
\bld
i I_{\bigtriangleup^{(b)}} \oint_\G \frac{dy}{2 \pi i y} \CA_{\f_a hh \bar{\f}_a}^{++} \CA_{\f_b \bar{\f}_b h}^{-} \CA_{\f_b \bar{\f}_b h}^{-} &= 0 \,,
\\ i I_{\bigtriangleup^{(b)}} \oint_\G \frac{dy}{2 \pi i y} \CA_{\f_a hh \bar{\f}_a}^{+-} \CA_{\f_b \bar{\f}_b h}^{-} \CA_{\f_b \bar{\f}_b h}^{+} &= \frac{3 G^2 \pi^2 m_a^2 m_b^3 (5\s^2 - 1)}{\sqrt{-t}} 
\,,
\eld \label{eq:b-top_results}
\el
to leading order in HCL. Summing over all possible intermediate graviton states is effectively equivalent to summing over these two terms and doubling the result. Summing over $a$-topology by exchanging $m_a \leftrightarrow m_b$, the $G^2$-order deflection angle is then given as
\bl
\bld
\vec{\vth}_2 &= \frac{\CE_{real}}{m_a m_b \sqrt{\s^2-1}} \left[ \frac{\partial \chi[\CA^{\bigtriangleup + \bigtriangledown}](b)}{\partial \vec{b}} \right]
\\ &= \frac{\CE_{real}}{(2 \sqrt{\s^2-1})^2} \frac{3\pi G^2 (m_a+m_b)}{\abs{b}^2} \left[- (5\s^2 - 1) \hat{y} \right] \,,
\eld \label{eq:2PM_scatt_ang}
\el
which, using the mapping of table \ref{tab:mapping} and making the approximation $m_a + m_b \simeq m_a = \CM_{\TN}$ yields a result inconsistent with geodesic computations \eqc{eq:2pmAnglevec} for non-zero $\vf$.
\bl
\vec{\vth}^{\rm Amp}_2 &=- \frac{3\pi \CM^{2}_{\TN}}{4}(5\CE^2 - \m^2)  \left(\frac{G}{J}\right)^{2}\hat{y} \,.
\el
There are a few notable points to these results.
\bn
\item The $(-t)^{-1/2}$ dependence solely comes from the scalar triangle integral $I_{\bigtriangleup^{(b)}}$. This means the residue integral $\oint \frac{dy}{y} \CA_4 \CA_3 \CA_3$ must scale as $(\b-1)^0$ to be consistent with the geodesic computations. The three-point amplitudes $\CA_3$ can be shown to scale as $(\b-1)^0$ through direct computations, therefore the condition can be refined to $(\b-1)^0$ scaling on the Compton amplitude $\CA_4$.
\item When the deflection angle \eqc{eq:2PM_scatt_ang} is evaluated using the na\"ive amplitude \eqc{eq:ComptonNaive}, only the helicity configurations invariant under the phase rotation contributes to the impulse. In other words, \emph{one-loop contributions to the impulse are indifferent to electric-magnetic rotations}. This is in stark contrast to the 2PM result of geodesic motion on a Taub-NUT background \eqc{eq:2pmAnglevec}, where $G^2$ order deflection angle had a dependence in the rotation angle $\vf$ through the combination $M = \CM_\TN \cos (2\vf)$.
\item The $b$-topology contribution \eqc{eq:b-top_results} scales as $\propto m_a^2 m_b^3$. Put differently, \emph{incorrectness of the Compton amplitude \eqc{eq:ComptonNaive} for NUT-charged particles is inconsequential for background-probe analysis} where NUT-charged particle acts as the background, because the contribution from the triangle topology where massive propagator on the background side has been cut dominates over the other triangle topology\footnote{If the Compton amplitude $\CA_{\f_a hh \bar{\f}_a}$ scales as $\propto m_a^3/m_b$, it would provide a loophole to this argument. Whether manifestation of Lorentz covariance breaking (the Dirac-Misner strings) can generate such a scaling will be left for future work.\label{fn:ComptonScaling}}.
\en

For completeness, possible resolutions for closing the gap between geodesic motion computations and amplitude results will be surveyed, to no avail.

\subsubsection{Compton hot-fix and impulse non-fix} \label{sec:ComptonFix}
The na\"ive amplitude \eqc{eq:ComptonNaive} is obviously wrong. As argued in the previous section, closing this loophole is unlikely to close the gap between geodesic computations and amplitude results. Nontheless, let us analyse the effects of na\"ive amplitude corrections to the impulse.

Although the na\"ive amplitudes \eqc{eq:ComptonNaive} do not satisfy the $t$-channel factorisation conditions, they do satisfy the factorisation conditions on the $s$- and $u$-channels. This implies that the difference between the correct amplitude and the na\"ive amplitude can only have poles on the $t$-channel,
\bl
\CA_{\text{correct}}-\CA_{\text{na\"ive}} &= \frac{f}{t} + \text{(polynomials)} \,, \label{eq:ComptonDiff}
\el
where the residue on the $t$-channel is proportional to the product of on-shell three-point amplitudes,
\bl
f \stackrel{t \to 0}{\propto} \CA_{\f \bar{\f} h} \CA_{hhh} \,,
\el
and the sum over exchanged $t$-channel graviton states has been suppressed. Although the exact expression for the difference \eqc{eq:ComptonDiff} is not available, the exactness of the expression is unimportant as leading HCL contributions dominate classical physics; when studying classical contributions from the $b$-topology cut of \fig{fig:b-top_cut}, the leading behaviour of the expression \eqc{eq:ComptonDiff} is all that matters\footnote{If triangle coefficient computation and HCL commutes; look at section \ref{sec:excitations} for a counterexample. However, all computations seem to imply that noncommutativity arises at subleading $(\b-1)$ orders.}.

The Mandelstam invariant $t$ for the Compton amplitude inside the $b$-topology cut is
\bl
(k_3+k_4)^2 = q^2 = - \frac{(\b-1)^2 m_b^2}{\b} \,,
\el
so the pole contributions dominate in the HCL $\b \to 1$. Therefore, the difference of the real amplitude and the na\"ive amplitude can be approximated as products of on-shell three-point amplitudes sitting on the $t$-channel pole.
\bl
\left. (\CA_{\text{correct}}-\CA_{\text{na\"ive}}) \right|_{\text{HCL}} \simeq \sum_{\eta = \pm} c_\eta \frac{\left. \CA_{\f \bar{\f} h}^{+\eta} \right|_{\text{HCL}} \left. \CA_{hhh}^{-\eta} \right|_{\text{HCL}}}{t} \,, \label{eq:CompDiffApprox}
\el
where the superscript $\eta$ denotes the helicity of the exchanged graviton and $c_\eta$ is the proportionality constant for that exchange contribution\footnote{This approximation relies on the observation that \emph{any} polynomial Lorentz invariants constructed from the momenta and spinors of external kinematics satisfying the little group weight constraints of gravitons vanish in the HCL. An obvious loophole is when spinors parametrising Dirac-Misner string degrees of freedom can be used to form Lorentz invariants; e.g. the invariant $[k_3 \xi]^4 [k_4 \xi]^4$ for $\CA^{++}$ will not vanish in the HCL when the spinor parametrising the Dirac-Misner string $\sket{\xi}$ is independent of HCL, unless $\sket{\xi} \propto \sket{\l}$.\label{fn:DMstringCompton}}. The on-shell three-point amplitudes can be approximated as
\bl
\bld
\left. \CA_{\f \bar{\f} h}^{\pm} \right|_{\text{HCL}} &= \left. \CA_{\f \bar{\f} h}^{\pm} (-p_1,p_1',-(k_3+k_4)) \right|_{\text{HCL}}
\eld \label{eq:3ptNUT}
\el
for the on-shell three-point amplitude of NUT-charged particle coupling to a graviton, and
\bl
\bld
\left. \CA_{hhh}^{\mp} \right|_{\text{HCL}} &= \lim_{\b \to 1} \left[ \CA_{hhh}^{\mp} (\left.(k_3+k_4)\right|_{\text{HCL}},-k_3,-k_4) \right]
\eld \label{eq:3ptGrav}
\el
for the on-shell three-graviton amplitude. While the amplitude \eqc{eq:3ptNUT} can be evaluated at the strict HCL, evaluating the amplitude \eqc{eq:3ptGrav} requires some care; the amplitude is first evaluated at momentum conservation violating kinematics where only $k_3+k_4$ is taken to the HCL, and then the resulting expression is taken to the HCL. The reason for taking such a convoluted sequence of limits is because the strict HCL is the collinear limit $k_3 \propto k_4$; $\CA_{hhh} = 0$ at HCL.

Direct computation of the HCL approximation \eqc{eq:CompDiffApprox} to the difference \eqc{eq:ComptonDiff} yields
\bl
\bld
\left. (\CA_{\text{correct}}^{++}-\CA_{\text{na\"ive}}^{++}) \right|_{\text{HCL}} &\simeq c_+^{++} \frac{\left. \CA_{\f \bar{\f} h}^{+} \right|_{\text{HCL}} \left. \CA_{hhh}^{-} \right|_{\text{HCL}}}{t} = - c_+^{++} \frac{u^2 y^2}{m_b^2} \,,
\\ \left. (\CA_{\text{correct}}^{+-}-\CA_{\text{na\"ive}}^{+-}) \right|_{\text{HCL}} &\simeq \sum_{\eta=\pm} c_\eta^{+-} \frac{\left. \CA_{\f \bar{\f} h}^{+\eta} \right|_{\text{HCL}} \left. \CA_{hhh}^{-\eta} \right|_{\text{HCL}}}{t} = - \frac{c_+^{+-} u^2 y^2 + c_-^{+-} v^2 y^2}{m_b^2} \,.
\eld \label{eq:CompDiffAppAns}
\el
Depending on helicity, the three-point amplitudes of the $b$-topology cut scales as $y^{\pm 2}$. This means augmentation to the Compton amplitude must scale as $y^0$ or $y^{\pm 4}$ to contribute to the triangle coefficient \eqc{eq:b-top_results}, but \eqc{eq:CompDiffAppAns} do not have the required scaling; fixing the na\"ive Compton amplitude will not fix the mismatch of scattering angle computations.

\subsubsection{Exciting internal structures and NUT-exciting results} \label{sec:excitations}
As argued in appendix \ref{app:monopole}, there is a strong tension in treating electrically charged and magnetically charged particles simultaneously as particles without internal structures, at least for abelian gauge theory. Although the same argument does not apply to gravity, the argument alludes to the possibility that NUT-charged particles cannot be structureless. This motivates the following question; will the disagreement between geodesic computations and amplitude computations be resolved when effects of NUT-charged particle's internal structures are taken into account?

One way to model internal structure excitations of the particle is to consider transmutation of quantum numbers such as mass or spin. This process can be captured by on-shell three-point amplitudes.
\bn
\item Equal masses; $\f \to \f' h$ where $m_\f = m_{\f'}$ and spin of $\f'$ can be $s'=0,1,2$.
\item Unequal masses; $\f \to \f' h$ where $m_\f \neq m_{\f'}$ and spin of $\f'$ is $s'=2$.
\en
The on-shell three-point amplitudes for the possibilities enumerated above can be kinematically determined as demonstrated in appendix \ref{app:3pt}. Introduction of these processes affect the $b$-topology cut, \fig{fig:b-top_cut}, through alterations in the Compton amplitude or in the pair of three-point amplitudes. The equal mass case $m_\f = m_{\f'} = m$ will be studied first, where cut momentum parametrisation \eqc{eq:cutmom} can still be used.

The modifications to the Compton amplitude from transmutation processes will always accompany $s$- or $u$-channel poles, therefore their primary effect appears on the four-particle cut; their effect on the three-particle cut is subleading in $(\b-1)$ compared to the na\"ive answer \eqc{eq:b-top_results}. As an explicit demonstration, when approximations to the modifications of the Compton amplitude similar to the previous section are adopted, computation of $b$-topology contributions to $\CA^{\bigtriangleup^{(b)}}$ from intermediate spin-1 (spin-2) particle contains additional powers of $(\b-1)$ and yields the scaling $\propto q^0$($\propto q^2$), which are contact interactions not affecting the long-distance physics\footnote{These results must be taken with a grain of salt; the scaling implies scalar integral coefficients are \emph{not} rational functions of external kinematics. For example, a proper computation using spin-2 exchange modification \eqc{eq:CompModS2} with $m_\f = m_{\f'}$ results in vanishing integral coefficients as can be seen in \eqc{eq:b-top_exc}.}. For this reason, only the modifications to the pair of three-point amplitudes appearing on the $b$-topology cut, \fig{fig:b-top_cut}, will be studied in detail for the equal mass case.

For the equal helicity exchange channel, the sum over intermediate states for the pair of three-point amplitude yields
\bl
\sum_{\f'} \CA_{\f \f' h}^{-}(k_3) \CA_{\f' \f h}^{-}(k_4) &= \left\{ \begin{aligned}
& \left(\frac{\g_0 \k m^2}{2}\right)^2 \frac{1}{y^4 \b^2} && s'=0
\\ & \left(\frac{\g_1 \k}{2}\right)^2 \frac{m^4 (\b-1)^2}{y^4 \b^3} && s'=1
\\ & \left(\frac{\g_2 \k}{2 m^2}\right)^2 \frac{m^8 (\b-1)^4}{y^4 \b^4} && s'=2
\end{aligned} \right.
\el
which implies $\oint \frac{dy}{y} \CA^{++} \CA^{-} \CA^{-} = 0$; the equal helicity exchange channel still does not contribute to the $b$-topology contribution $\CA^{\bigtriangleup^{(b)}}$. On the other hand, for the mixed helicity exchange channel the same computations lead to
\bl
\sum_{\f'} \CA_{\f \f' h}^{-}(k_3) \CA_{\f' \f h}^{+}(k_4) &= \left\{ \begin{aligned}
& \left(\frac{\g_0 \k m^2}{2}\right)^2 \b^2 && s'=0
\\ & \left(\frac{\g_1 \k}{2}\right)^2 m^4 (\b-1)^2 \b && s'=1
\\ & \left(\frac{\g_2 \k}{2 m^2}\right)^2 m^8 (\b-1)^4 && s'=2
\end{aligned} \right. \label{eq:eqmasstransmute}
\el
which implies that $\g^2 (\b-1)^{2s'} \simeq \g^2 \left(\frac{-t}{m_b^2}\right)^{s'}$ factor is attached to the original $b$-topology contribution $\CA^{\bigtriangleup^{(b)}}$. An interesting observation point is that for $s'=2$ transmutations, the modification to the $b$-topology contribution $\D \CA^{\bigtriangleup^{(b)}}$ scales as $\propto \abs{q}^3 \sim r^{-6}$, which is known to be the $1/r$ dependence for tidal effect corrections~\cite{Cheung:2020sdj}. While the resulting $\s$-dependence for $m_\f = m_{\f'}$ case is inconsistent with the results of ref.\cite{Cheung:2020sdj}, the conclusions are different for the case of $m_\f \neq m_{\f'}$.

For unequal mass case $m_{\f'} \neq m_\f$, the Compton amplitudes are modified by inclusion of $\f'$ exchange on the $s$- and $u$-channels as
\bl
\bld
\D \CA_{\f hh \bar{\f}}^{++} &= - \frac{\g'^2 (\k/2)^2 m_{\f'}^4 [23]^4}{m_\f^4} \left( \frac{1}{s-m_{\f'}^2} + \frac{1}{u-m_{\f'}^2} \right) \,,
\\ \D \CA_{\f hh \bar{\f}}^{+-} &= - \frac{\g'^2 (\k/2)^2 \sbra{2} p_1 \ket{3}^4}{m_\f^4} \left( \frac{1}{s-m_{\f'}^2} + \frac{1}{u-m_{\f'}^2} \right) \,,
\eld \label{eq:CompModS2}
\el
where the three-point amplitudes \eqc{eq:uneqmass_transmute} have been glued on the channels as in ref.\cite{Arkani-Hamed:2017jhn} to arrive at the results. The absolute square of the coupling $\abs{\g'}^2$ is related to the characteristic decay rate of resonance excitations and $m_{\f'}^2 - m_\f^2 \simeq 2 m_\f \D m$ is related to the characteristic frequency of the resonance excitations $\hbar \omega = \D m$.

Interestingly, the order of taking the HCL and effecting the $\oint \frac{dy}{y}$ integral matters. When constituent tree amplitudes are taken to the HCL before $\oint \frac{dy}{y}$ integration, the modifications to $b$-topology contribution are computed as
\bl
\bld
i I_{\bigtriangleup^{(b)}} \oint_\G \frac{dy}{2 \pi i y} \left[ \D \CA_{\f_a hh \bar{\f}_a}^{++} \right] \CA_{\f_b \bar{\f}_b h}^{-} \CA_{\f_b \bar{\f}_b h}^{-} &= G^2 \pi^2 m_b^3 \left[ \frac{4 \g_a'^2 (m_{a'}/m_a)^4}{m_{a'}^2 - m_a^2} \right] \sqrt{-t}^3 \,,
\\ i I_{\bigtriangleup^{(b)}} \oint_\G \frac{dy}{2 \pi i y} \left[ \D \CA_{\f_a hh \bar{\f}_a}^{+-} \right] \CA_{\f_b \bar{\f}_b h}^{-} \CA_{\f_b \bar{\f}_b h}^{+} &= G^2 \pi^2 m_b^3 \left[ \frac{\g_a'^2 (3 - 30\s^2 + 35 \s^4)}{2(m_{a'}^2 - m_a^2)} \right] \sqrt{-t}^3
\,,
\eld \label{eq:b-top_tidal}
\el
which scales as $\propto \abs{q}^3 \sim r^{-6}$ and can be recast as
\bl
\bld
\D \CA^{\bigtriangleup^{(b)}} &= G^2 \pi^2 m_b^3 \sqrt{-t}^3 \left[ \g_a'^2 \frac{8 (m_{a'}/m_a)^4 + (3 - 30\s^2 + 35 \s^4)}{m_{a'}^2 - m_a^2} \right]
\\ &\simeq G^2 \pi^2 m_b^3 \sqrt{-t}^3 \left[ \g_a'^2 \frac{(11 - 30\s^2 + 35 \s^4) + 32 (\D m_a / m_a)}{2 m_a \D m_a} \right] \,,
\eld \label{eq:tidal_tot}
\el
where lowest order expansion in $\D m_a = m_{a'} - m_a$ has been performed on the last line. This has an interpretation as tidal effects; comparing with eq.(4) of ref.\cite{Cheung:2020sdj}, the tidal response coefficients $\l$ and $\eta$ can be related to $\g'$ and $\D m$ by\footnote{If there are more than one resonance modes, the relation becomes a sum over different resonance modes. A quantum mechanical version of the relation has been considered in ref.\cite{Brustein:2020tpg}.}
\bl
\l = \frac{4 \g'^2}{m^2} \,,\, \eta = \frac{16 \g'^2}{m \D m} \,, \label{eq:transmute2tidal}
\el
which in turn relates them to the worldline tidal operator coefficients through eq.(12) of the same reference.
\bl
\s^{(2)} = - \frac{12 \g'^2}{m^3} \,,\, \m^{(2)} = \frac{16 \g'^2}{m^2 \D m} + \frac{96 \g'^2}{3 m^3} \,. \label{eq:transmute2tidal2}
\el
The above equation relates the parameters of dissipation through resonance excitations to that of tidal responses, which is reminiscent of Kramers-Kronig relations.

On the other hand, performing the $\oint \frac{dy}{y}$ integration and then taking the HCL results in
\bl
\bld
i I_{\bigtriangleup^{(b)}} \oint_\G \frac{dy}{2 \pi i y} \left[ \D \CA_{\f_a hh \bar{\f}_a}^{++} \right] \CA_{\f_b \bar{\f}_b h}^{-} \CA_{\f_b \bar{\f}_b h}^{-} &= 0 \,,
\\ i I_{\bigtriangleup^{(b)}} \oint_\G \frac{dy}{2 \pi i y} \left[ \D \CA_{\f_a hh \bar{\f}_a}^{+-} \right] \CA_{\f_b \bar{\f}_b h}^{-} \CA_{\f_b \bar{\f}_b h}^{+} &= - \frac{4 \g_a'^2 G^2 \pi^2 m_a^2 m_b^3}{\sqrt{-t}} \left[ \frac{m_{a'}^2 - m_a^2}{m_a^2} \right]^3
\,,
\eld \label{eq:b-top_exc}
\el
which has the wanted scaling $\propto \abs{q}^{-1} \sim r^{-2}$. However \eqc{eq:b-top_exc} cannot resolve the discrepancy between geodesic computations and amplitude computations as the mass scaling is $\propto m_a^2 m_b^3$, similar to na\"ive Compton amplitude impulse computation \eqc{eq:b-top_results}; if the particle $a$ having internal structures is considered as the background, then this contribution is suppressed by the mass ratio $m_b/m_a$ compared to the $a$-topology contribution.

When the massive propagator of the $\f'$ particle is cut as in \fig{fig:b-top_cut} with the condition $m_{\f'} \neq m_\f$, the leading small $q^2$ contributions to $\D \CA^{\bigtriangleup^{(b)}}$ are computed as\footnote{The cut momentum parametrisations \eqc{eq:cutmom} are no longer valid, and the new cut conditions $k_3^2 = k_4^2 = L^2 - m_{b'}^2 = 0$ were solved to arrive at the result \eqc{eq:uneqmasstransmute}.}
\bl
\bld
i I_{\bigtriangleup^{(b)}} \oint_\G \frac{dy}{2 \pi i y} \CA_{\f_a hh \bar{\f}_a}^{++} \CA_{\f_b \f'_b h}^{-} \CA_{\f_b \f'_b h}^{-} &= 0 \,,
\\ i I_{\bigtriangleup^{(b)}} \oint_\G \frac{dy}{2 \pi i y} \CA_{\f_a hh \bar{\f}_a}^{+-} \CA_{\f_b \f'_b h}^{-} \CA_{\f_b \f'_b h}^{+} &= \frac{\g_b'^2 G^2 \pi^2 m_a^2 m_b^5 (3 \s^2 - 1)}{\sqrt{-t}^3} \left[ \frac{m_b}{m_{b'}} \right] \left[ \frac{m_{b'}^2-m_b^2}{m_b^2} \right]^6
\,.
\eld \label{eq:uneqmasstransmute}
\el
The interpretation of this result is unclear: Although the overall coefficient $(\D m_{b} / m_b)^6$ highly suppresses the result, the above scales as $\abs{q}^{-3} \sim \log r$ which actually \emph{grows} at long distances. If the limit $m_{b'} \to m_b$ is taken before the limit $q^2 \to 0$ then the resulting cut computation reduces to \eqc{eq:eqmasstransmute} instead, therefore it is unlikely that computations leading to the above result are wrong. How to correctly interpret this result will be left for a future work.

In conclusion, exciting the NUT charges, modelled as transmutations to distinct particle species, cannot close the gap between geodesic computations of general relativity and amplitude computations at one-loop. While the results are not exciting from the viewpoint of original motivations, the relation \eqc{eq:transmute2tidal} between internal excitation parameters and tidal deformation parameters obtained from \eqc{eq:tidal_tot} is an interesting by-product of the analysis that calls for future studies.

\section{Conclusion} \label{sec:conc}

In this study the classical impulse at the second post-Minkowskian order ($G^2$) was analysed from Hamilton-Jacobi analysis of general relativity on Taub-NUT spacetime and quantum field theory one-loop amplitude calculations. Although the computations agreed at 1PM, the computations were shown to disagree at 2PM. Possible modifications to one-loop amplitude computations that could resolve the disagreement had been analysed, but all considered scenarios had failed to close the gap. 

There are a few loopholes to the amplitude analysis, one being possible modifications to the amplitudes due to presence of Dirac-Misner strings. As mentioned in footnote \ref{fn:ComptonScaling}, the argument that only the $b$-topology cut of \fig{fig:b-top_cut} with $m_b \gg m_a$ is relevant for 2PM scattering angle analysis in the background-probe limit is based on the expectation that the Compton amplitude $\CA_{\f_b h h \bar{\f}_b}$, which appears in the $a$-topology cut, knows nothing about the mass of particle $a$. If the degrees of freedom parametrising the Dirac-Misner string attached to particle $b$ somehow knows about $m_a$ then the argument based on \eqc{eq:b-top_results} that $a$-topology contribution is negligible would break down. The approximations to the difference of correct and na\"ive Compton amplitude \eqc{eq:CompDiffApprox} also contains a similar loophole as explained in footnote \ref{fn:DMstringCompton}.

It is also possible that application of unitarity methods found in the literature need generalisations when magnetic charges are involved. For example, it is unclear how pairwise little group introduced in ref.\cite{Csaki:2020inw}, which demands introduction of a new quantum number (pairwise helicity) for each dyonic pair of particles, can be applied to unitarity methods. When unitarity methods are applied to loop effects for electric charge-monopole scattering, the cuts can be performed in a way that all resulting tree diagrams only involve one type of charges. This results in an awkward situation where none of the constituent subdiagrams require pairwise helicity but the diagram as a whole does.

It would be an interesting exercise to ponder the ramifications of the relation \eqc{eq:transmute2tidal} and its generalisations. Black holes of general relativity in four space-time dimensions are known to be inert to tidal forces, expressed in the form of vanishing Love numbers~\cite{Porto:2016zng}. The relation implies that vanishing of tidal Love numbers poses a constraint on quantum mechanical excitations of the black hole \cite{Kim:2020dif}. We also remark that the relation is similar in spirit to the Fermi four-fermion interactions being completed to weak interactions of the standard model.

\acknowledgments
We would like to thank to Yu-Tin Huang for technical discussions, Seok Kim for helpful suggestions, and our Ph.D. advisors, Sangmin Lee and Nakwoo Kim, for feedbacks.
JK and MS also thank for the hospitality of Kyung Hee University and Seoul National University during their respective visits. JK was supported in part by the Science and Technology Facilities Council (STFC) Consolidated Grant ST/T000686/1 \emph{“Amplitudes, Strings and Duality”}, and in part by National Research Foundation (NRF) of Korea grant NRF-2019R1A2C2084608. MS was supported in a part by National Research Foundation (NRF) of Korea grant NRF-2019R1A2C2004880 and NRF-2018R1A2A3074631, and a scholarship from Hyundai Motor Chung Mong-Koo Foundation.  

\appendix
\section{The elementarity tension between electric charges and monopoles} \label{app:monopole}
It is well known that Lagrangian field theory description of a system including electrically charged particles and magnetically charged particles as elementary degrees of freedom in its spectrum simultaneously can only be attained at the price of (1) manifest Lorentz invariance or (2) manifest local description~\cite{Hagen:1965zz,Zwanziger:1970hk,Schwarz:1993vs,Dirac:1948um,Schwinger:1966nj}. Nevertheless, the following question may be asked; \emph{is it eligible to describe electrically charged particles and magnetically charged particles as "elementary particles" in an effective theory whose cut-off is much higher than the particles' masses?} Here, "elementary particle" refers to particles whose internal structures can be ignored and regarded as point particles. For example, in sufficiently low energy experiments protons and pions can be regarded as "elementary particles".

A simple argument shows this is unlikely. The \emph{classical radius} $r_c$ is the scale where point particle description of the body breaks down in the classical theory, and the \emph{Compton wavelength} $\l_C$ is the scale where classical description of the body breaks down and quantum theory must be introduced. If $r_c>\l_C$, there is an intermediate length scale $L$, $r_c>L>\l_C$, such that its non-point-particle-like structure must become manifest because it cannot be clouded by quantum effects. Therefore, \emph{for a particle to be described as an "elementary particle" its Compton wavelength must be greater than its classical radius; $\l_C > r_c$}. A slightly stronger condition $\l_C \gg r_c$ cannot be imposed on electric and magnetic charges simultaneously due to Dirac-Schwinger-Zwanziger quantisation conditions.

Ignoring $O(1)$ numerical coefficients, the classical radius and the Compton wavelength of an electrically charged particle of mass $m$ and unit charge $e$ is given as;
\bg
r_c \simeq \frac{e^2}{m c^2} \,,\, \l_C \simeq \frac{\hbar}{m c} \;\Rightarrow \; \frac{r_c}{\l_C} \simeq \frac{e^2}{\hbar c} = \a_{e} \,.
\eg
Likewise, the ratio for a magnetically charged particle of unit charge is given as $r_c / \l_C \simeq g^2/\hbar c = \a_{g}$. The least restrictive Dirac-Schwinger-Zwanziger quantisation condition demands $eg \simeq O(1)$, which translates to $\a_e \a_g \simeq O(1)$; the "elementary particle" conditions $\a_e \ll 1$ and $\a_g \ll 1$ are not compatible with the quantisation conditions, unless the effective theory cut-off scale $\L_{\text{EFT}}$ screens the classical radius and allows "elementary particle" descriptions. In the latter case, the condition $\L_{\text{EFT}} < m$ must be satisfied by the heavier of the charges\footnote{Essentially the same argument appears in magnetic version of the weak gravity conjecture~\cite{ArkaniHamed:2006dz}.}.

Note that the "classical radius" for gravitational interactions is also a scale where Newtonian gravity becomes inadequate; the radius can be defined as the scale where Newtonian gravitational self-energy becomes comparable to its rest mass. Up to $O(1)$ numerical coefficients,
\bl
m c^2 \simeq \frac{G m^2}{r_{c}} \; \Rightarrow \; r_c \simeq \frac{G m}{c^2} \,,
\el
which is the scale of the body's Schwarzschild radius; general relativity must be introduced at this scale.

\section{All on-shell three-point amplitudes corresponding to transmutation} \label{app:3pt}
All kinematicly allowed gauge-invariant on-shell three-point amplitudes of the type $\CA_{\f \f' h}$ are enumerated in this appendix, which are special cases of the classification initiated in ref.\cite{Arkani-Hamed:2017jhn}. The massive spinor-helicity variables of ref.\cite{Arkani-Hamed:2017jhn} is used for massive spinning particles and conventions for the variables follow that of ref.\cite{Chung:2018kqs}.

For sake of simplicity, only the coupling to positive helicity graviton will be considered. The coupling to negative helicity can be achieved by exchanging\footnote{Couplings are assumed to be real.} square and angle spinor brackets $\la pq \ra \leftrightarrow [pq]$. The little group constraint on $\CA_{\f \f' h}^{+}$ for the graviton leg is
\bl
\sket{3}^{4-n} \ket{3}^{-n} \subset \CA_{\f \f' h}^{+} \,,
\el
where $\sket{3}$ and $\ket{3}$ are spinors for the graviton leg and the exponent shows how many times each spinor appears on the numerator/denominator of the amplitude.

\subsection{Equal mass}
When $m_\f = m_{\f'} = m$, the kinematic factor $x$ of \eqc{eq:xdef} introduced in ref.\cite{Arkani-Hamed:2017jhn} can be used to provide additional little group weight for the graviton leg. Three possibilities can be considered.
\bl
\CA_{\f \f' h}^{+} = \left\{ \begin{aligned}
& \g_0 \frac{\k m^2}{2} x^2 && s' = 0
\\ & \g_1 \frac{\k}{2} x [3 \mathbf{2}]^2 && s' = 1
\\ & \g_2 \frac{\k}{2m^2} [3 \mathbf{2}]^4 && s' = 2
\end{aligned} \right.
\el
In the above expression, the transmuted particle $\f'$ has been given the leg number 2 and the couplings $\g_i$'s have been scaled to be dimensionless using $\k$ and $m$.

\subsection{Unequal mass}
The kinematic factor $x$ of \eqc{eq:xdef} is gauge-invariant only if $m_\f = m_{\f'}$, where gauge-invariance denotes invariance of the expression under the gauge transform $\ve^\m (p_3) \sim \ve^\m (p_3) + c p_3^\m$. Gauge invariance enforces the condition $p_1 \cdot p_3 = 0$, which in turn enforces $p_1^2 = p_2^2$; the existence of $x$-factor demands\footnote{$m>0$ is implicitly assumed.} $m_\f = m_{\f'}$. This means four more spinors must be introduced to form a Lorentz invariant from four $\sket{3}^{\dot\a}$ spinors of the graviton leg. The only possibility is to introduce at least four spinors of the $\f'$ leg, meaning the transmuted particle $\f'$ must have spin $s' \ge 2$. Saturating the bound $s'=2$ yields the amplitude
\bl
\CA_{\f \f' h}^{+} = \g' \frac{\k}{2m_\f^2} [3 \mathbf{2}]^4 \,. \label{eq:uneqmass_transmute}
\el
This result is consistent with angular momentum conservation and can be understood as a gravitational analogue of the selection rule $\D L = \pm 1$ for dipole transitions of the hydrogen atom; when $m_\f \neq m_\f'$, a real process of heavier of the two decaying into the lighter of the two by emitting a graviton is allowed.

Assuming $m_\f > m_\f'$, the decay process $\f \to \f'h$ can be analysed at the rest frame of $\f$. In this frame the total angular momentum of the initial state $\f$ is zero, which must be conserved after the decay. Since orbital angular momentum of the final state $\f' h$ cannot contribute to the total angular momentum directed along the direction of the emitted graviton, the spin component along this direction for the decay products must sum up to zero. The graviton has spin 2 along this direction, so the decay product $\f'$ also must have spin 2 along this direction with opposite signs. Extension of the analysis for the case $m_\f < m_\f'$ is straightforward.



\section{Dyonic generalisation of effective one-body formalism}
In Newtonian mechanics, the two-body problem can be reduced to a one-body problem using reduced mass. The \emph{effective one-body} (EOB) formalism is an attempt to generalise the reduction to relativistic systems, where the two-body motion is mapped to an effective one-body geodesic motion on a background metric~\cite{Buonanno:1998gg}. The formalism often reduces to finding a map between the real energy $\CE_{\rm real}$ of the two-body system and the effective energy $\CE_{\eff}$ of the one-body geodesic motion, which smoothly interpolates between background-probe limit and comparable mass case. In the extreme mass-ratio limit the effective metric often reduces to known black hole solutions. To find the energy map, one first sets the action-angle variables of the real two-body system and the effective one-body system to be the same, and then inverts the relation between energy and action-angle variables to obtain a relation between the energy of the real system and the energy of the effective system.

We will attempt a generalisation of effective one-body formalism for scattering states~\cite{Damour:2016gwp} to cases involving "magnetic mass", using amplitude computations as the input for real dynamics. The dynamics obtained from amplitudes defies an interpretation as the motion on a Taub-NUT background with backreaction of the probe particle taken into account, as the extreme mass-ratio limit is not the geodesic motion on a Taub-NUT background. Therefore, an interesting problem would be to find the effective background metric the dynamics corresponds to. We adopt the usual \emph{total mass} $\CM_{total}$ and \emph{effective mass} $\m$ relations for inertial masses.
\beq
\CM^2\equiv M^2+N^2,&&\CM_{total}=\CM_{a}+\CM_{b},,&&\m=\frac{\CM_a \CM_b }{\CM_{total}}\,.
\eeq
The electric and magnetic masses are given as
\beq
\cos(2\vf) =\frac{M}{\mathcal{M}_{\TN}}\leftrightarrow\frac{M_{\eff}}{\mathcal{M}_{total}} 
,&&\sin(2\vf)=\frac{N}{\mathcal{M}_{\TN}}\leftrightarrow\frac{N_{\eff}}{\mathcal{M}_{total}},
\eeq
\bl
b\abs{\vec{p}_{\COM}} &= \frac{m_a m_b b\sqrt{\s^2 - 1}}{\sqrt{s}}=J\Leftrightarrow b\sqrt{\s^2-1}=\frac{J\CE_{real}}{m_a m_b} \,.
\el
If both particles undergo dyonic rotation, only the relative rotation angle $\vf=\vf_a-\vf_b$ affects the dynamics. The angles can be written using the masses as given below.
\beq
\cos{2\vf_{a}}=\frac{M_a}{\CM_{a}},&&\cos{2\vf_{b}}=\frac{M_b}{\CM_{b}},&&\cos{2\vf}=\frac{M_{\eff}}{\CM_{total}}=\frac{M_{a}M_{b}+N_a N_b}{\CM_{a}\CM_{b}},
\\
\sin{2\vf_{a}}=\frac{N_a}{\CM_{a}},&&
\sin{2\vf_{b}}=\frac{N_b}{\CM_{b}},&&
\sin{2\vf}=\frac{N_{\eff}}{\CM_{total}}=\frac{M_{a}N_{b}-M_{b}N_{a}}{\CM_{a}\CM_{b}}\,.
\eeq
The effective dynamics will depend on the effective electric(magnetic) mass $M_{\eff}$($N_{\eff}$), which can be written using the reduced mass $\m$.
\beq
M_{\eff}=\frac{M_{a}M_{b}+N_a N_b}{\m},&&N_{\eff}=\frac{M_a N_b - M_b N_a}{\m}\,.
\eeq
For the rest of this section, we drop the subscripts on $M_{\eff}$ and $N_{\eff}$ for simplicity.

\subsection{The deflection angle of effective dynamics}

Let us consider an effective metric\footnote{$\b$ denotes deformation parameters.} which generalises the deformed Schwarzschild metric in ref.\cite{Damour:2016gwp} up to second Post-Minkowskian order.
\beq
ds^{2}_{\eff}&=-f_{1}dt^2+f_{2}dr^2+(r^2+G^2 N^2)d\W_{2},
\eeq
where
\beq
f_1=1-\frac{2 G M}{r}-\frac{2 G^2 N^2}{r^2},&&f_{2}(\b)=1+\b_{1}\frac{2 G M}{r}+\b_{2}\frac{(2GM)^2}{r^2}+\b_{3}\frac{2(GN)^2}{r^2}.
\eeq
The corresponding effective Hamilton-Jacobi equation on the cone is
\beq
-f_{1}^{-1}\CE_{\eff}^{2}+\frac{J^{2}_{\vf}}{(r^{2}+N^{2})}+f_{2}^{-1}P(r)^{2}=-\m^{2},
\eeq
or, equivalently,
\beq
P_{\pm}(r)=\pm\left[\frac{f_2}{f_1}\CE_{\eff}^{2}-f_{2}\m^{2}-f_{2}\frac{J^{2}_{\vf}}{(r^{2}+N^{2})}\right]^{\half}.
\eeq
The effective deflection angle can be obtained as
\beq
\vth_{\eff}&=2 \sqrt{\frac{M^2 \left(\left(\beta _1+1\right) \CE_{\eff}^2-\beta _1 \mu ^2\right)^2}{\CE_{\eff}^2-\mu ^2}+4 N^2 \CE_{\eff}^2}\left(\frac{ G}{J}\right)\nn
\\
&\qquad+\pi\frac{  M \left(\left(\beta _1+1\right) \CE_{\eff}^2-\beta _1 \mu ^2\right)}{4 \sqrt{M^2 \left(\left(\beta _1+1\right) \CE_{\eff}^2-\beta _1 \mu ^2\right){}^2+4 N^2 \CE_{\eff}^2 \left(\CE_{\eff}^2-\mu ^2\right)}}\nn
\\
&\qquad\qquad\qquad\times\Bigl[M^2 \left(\left(\beta _1^2-4 \beta _1-4 \beta _2-8\right) \CE_{\eff}^2-\left(\beta_1^2 -4 \beta _2\right) \mu ^2\right)\nn
\\
&\qquad\qquad\qquad\qquad\qquad\qquad+N^2 \left(\left(2 \beta_3+1\right) \mu^2+\left(3-2 \beta_3\right) \CE_{\eff}^2\right)\Bigr]\left(\frac{G}{J}\right)^{2}.
\eeq

\subsection{The 1PM mapping}
Since the action-angle variables are matched in the effective one-body formalism, we set the deflection angles equal; $\vth_{\eff} = \vth_{\rm real}$. This results in
\beq
\sqrt{\frac{1}{( \m^2\s^2-\m^2)} (2\m^2 \s^2 - \m^2)^{2} M^2  +4\m^2 \s^{2} N^2}=\sqrt{\frac{M^2 \left(\left(\beta _1+1\right) \mathcal{E}_{\eff}^2-\beta _1 \mu ^2\right){}^2}{\mathcal{E}_{\eff}^2-\mu ^2}+4 N^2 \mathcal{E}_{\eff}^2}.
\eeq
Adopting new dimensionless variables, $u_r^2=(\s^2-1)$ and $u_{e}^2=\left(\CE^2 / \m^2-1\right)$, the relation is recast into
\beq
g_{1,r}(u_r)M^2+g_{2,r}(u_r)4N^2=g_{1,e}(u_e)M^2+g_{2,e}(u_e)4N^2,
\eeq
where
\beq
\begin{split}
g_{1,r}(u_r)&=(2u_{r}+u_{r}^{-1})^{2},
\\
g_{2,r}(u_r)&=(u_{r}^{2}+1),
\end{split}
&&
\begin{split}
g_{1,e}(u_e)&=(\left(\beta _1+1\right)u_{e}+u_{e}^{-1} )^2,
 \\
g_{2,e}(u_e)&=(u_{e}^2+1).
\end{split}
\eeq
The parameter $\b_1$ can be determined as in ref.\cite{Damour:2016gwp}, where the criteria for fixing $\b_1$ is given as 
\beq
\begin{split}
g_{1,r}(u_r)\sim g_{1,e}(u_e)&\Rightarrow u^{\min(g_{2,r})}_{r}=\frac{1}{\sqrt{2}},u^{\min(g_{2,e})}_{e}=\frac{1}{\sqrt{\b_1+1}},
\\
g_{2,r}(u_r)\sim g_{2,e}(u_e)&\Rightarrow u^{\min(g_{2,r})}_{r}=0,\ph{aa}u^{\min(g_{2,e})}_{e}=0.
\end{split}
\eeq
Solving for the relations yields
\beq
\b_1=1,&&\s^{2}=\frac{\CE_{\eff}^2}{\m^2}=\frac{\CE_{\rm real}^2-\CM_{a}^{2}-\CM_{b}^{2} }{2\CM_{a}\CM_{b}}.
\eeq
Combining everything, we arrive at the effective one-body energy mapping at 1PM order.
\beq
\CE_{\eff}=\frac{\CE_{\rm real}^2-\CM_{a}^{2}-\CM_{b}^{2} }{2\CM_{total}}.\label{eq:eob1pm}
\eeq
Taking the limit $N_i \to 0$ reduces the above relation to the known answers of ref.\cite{Damour:2016gwp}.

\subsection{The 2PM mapping}
Using $\b_1=1$ from first order matching, the second order deflection angle with undetermined parameters $\{\b\}$ of the effective metric is given as 
\beq
\left|\vec{\vth}^{\eff}_{G^2}\right|&=-\pi\frac{  M \m^2  \left(2 \frac{\CE_{\eff}^2}{\mu ^2}-1\right)}{4 \sqrt{M^2+4\CM_{\TN}^2\frac{\CE_{\eff}^2}{\m^2}(\frac{\CE_{\eff}^2}{\m^2}-1)}}\nn
\\
&\qquad\times \Bigl[M^2 \left(\left(14+4 \beta _2-2 \beta _3\right) \frac{\CE_{\eff}^2}{\mu ^2}+\left(2-4 \beta _2+2 \beta _3\right)\right)
\nn
\\
&\qquad\qquad-\CM_{\TN}^2 \left(\left(3-2 \beta _3\right) \frac{\CE_{\eff}^2}{\mu ^2}+\left(2 \beta _3+1\right) \right)\Bigr]\left(\frac{ G}{J}\right)^{2} \,.\label{eq:eTN2pm}
\eeq
In the Schwarzschild limit, the effective deflection angle reduces to 
\beq
\abs{\vth_{Sch}}^{\eff}_{G^2}&=-\frac{\p \CM_{total}^2 \m^2 }{4}\Bigl[\left(\left( 4 \beta _2+11\right) \frac{\CE_{\eff}^2}{\mu ^2}+\left(1-4 \beta _2\right) \right)\Bigr]\left(\frac{ G}{J}\right)^{2}.\label{eq:eSch2pm}
\eeq
This is in contrast to the second order result from amplitude computations, which does not depend on the phase $\vf$ but only on the inertial masses.
\beq
\abs{\vth}^{\rm Amp}_2&=-\frac{\p \CM_{total}^{2}\m^2}{4}(15\s^2 - 3)\left(\frac{ G}{J}\right)^{2}.\label{eq:2pmAmp}
\eeq
Apparently (\ref{eq:2pmAmp}) cannot be reproduced from (\ref{eq:eTN2pm}) by adjusting the $\b_2$ and $\b_3$, but can be matched to (\ref{eq:eSch2pm}) under absence of NUT charges. It seems our expectation of finding a consistent effective metric of EOB formalism that matches amplitude computations cannot be met by the generalisation considered.

Using the same dimensionless variables, $u$, we can directly compare the both and determine $\b_2$ for the $N\to0$ limit. 
\beq
15u_{r}^{2}+12=(4\b_2+11)u^{2}_{e}+12&&\Leftrightarrow&& g_{3,r}(u_r)=g_{3,e}(u_e;\b_{2}).
\eeq
It is suffice to solve for the zeros of the functions to be the same, $u_{r,0}=u_{e,0}$.
\beq
u_{r,0}^2=-\frac{12}{15}=u_{e,0}^{2}=-\frac{12}{4\b_2+11}\Rightarrow\b_{2}=1.
\eeq
The two solutions, $\b_1=1$ and $\b_2=1$, correspond to the second order post-Minkowskian expansion of the Schwarzschild solution. This gives us the same energy map \eqc{eq:eob1pm} obtained from first order analysis. 


\bibliography{ref.bib}

\providecommand{\href}[2]{#2}\begingroup\raggedright\begin{thebibliography}{10}

\bibitem{Huang:2019cja}
Y.-T. Huang, U.~Kol and D.~O'Connell, \emph{{Double copy of electric-magnetic
  duality}}, \href{http://dx.doi.org/10.1103/PhysRevD.102.046005}{\emph{Phys.
  Rev. D} {\bf 102} (2020) 046005},
  [\href{http://arxiv.org/abs/1911.06318}{{\tt 1911.06318}}].

\bibitem{Emond:2020lwi}
W.~T. Emond, Y.-t. Huang, U.~Kol, N.~Moynihan and D.~O'Connell,
  \emph{{Amplitudes from Coulomb to Kerr-Taub-NUT}},
  \href{http://arxiv.org/abs/2010.07861}{{\tt 2010.07861}}.

\bibitem{Taub:1950ez}
A.~Taub, \emph{{Empty space-times admitting a three parameter group of
  motions}}, \href{http://dx.doi.org/10.2307/1969567}{\emph{Annals Math.} {\bf
  53} (1951) 472--490}.

\bibitem{Newman:1963yy}
E.~Newman, L.~Tamburino and T.~Unti, \emph{{Empty space generalization of the
  Schwarzschild metric}}, \href{http://dx.doi.org/10.1063/1.1704018}{\emph{J.
  Math. Phys.} {\bf 4} (1963) 915}.

\bibitem{Misner:1963fr}
C.~W. Misner, \emph{{The Flatter regions of Newman, Unti and Tamburino's
  generalized Schwarzschild space}},
  \href{http://dx.doi.org/10.1063/1.1704019}{\emph{J. Math. Phys.} {\bf 4}
  (1963) 924--938}.

\bibitem{Ramaswamy81}
S.~Ramaswamy and A.~Sen, \emph{Dual‐mass in general relativity},
  \href{http://dx.doi.org/10.1063/1.524839}{\emph{Journal of Mathematical
  Physics} {\bf 22} (1981) 2612--2619}.

\bibitem{Ramaswamy:1986kf}
S.~Ramaswamy and A.~Sen, \emph{{COMMENT ON `GRAVITOMAGNETIC POLE AND MASS
  QUANTIZATION.'}},
  \href{http://dx.doi.org/10.1103/PhysRevLett.57.1088}{\emph{Phys. Rev. Lett.}
  {\bf 57} (1986) 1088}.

\bibitem{Kol:2020zth}
U.~Kol, \emph{{Dual Komar Mass, Torsion and Riemann-Cartan Manifolds}},
  \href{http://arxiv.org/abs/2010.07887}{{\tt 2010.07887}}.

\bibitem{Carter:1968ks}
B.~Carter, \emph{{Hamilton-Jacobi and Schrodinger separable solutions of
  Einstein's equations}},
  \href{http://dx.doi.org/10.1007/BF03399503}{\emph{Commun. Math. Phys.} {\bf
  10} (1968) 280--310}.

\bibitem{bonnor69}
W.~B. Bonnor, \emph{A new interpretation of the nut metric in general
  relativity},
  \href{http://dx.doi.org/10.1017/S0305004100044807}{\emph{Mathematical
  Proceedings of the Cambridge Philosophical Society} {\bf 66} (1969)
  145–151}.

\bibitem{Dowker74}
J.~S. {Dowker}, \emph{{The nut solution as a gravitational dyon}},
  \href{http://dx.doi.org/10.1007/BF02451402}{\emph{General Relativity and
  Gravitation} {\bf 5} (Oct., 1974) 603--613}.

\bibitem{Alawadhi:2019urr}
R.~Alawadhi, D.~S. Berman, B.~Spence and D.~Peinador~Veiga, \emph{{S-duality
  and the double copy}},
  \href{http://dx.doi.org/10.1007/JHEP03(2020)059}{\emph{JHEP} {\bf 03} (2020)
  059}, [\href{http://arxiv.org/abs/1911.06797}{{\tt 1911.06797}}].

\bibitem{Kol:2019nkc}
U.~Kol and M.~Porrati, \emph{{Properties of Dual Supertranslation Charges in
  Asymptotically Flat Spacetimes}},
  \href{http://dx.doi.org/10.1103/PhysRevD.100.046019}{\emph{Phys. Rev. D} {\bf
  100} (2019) 046019}, [\href{http://arxiv.org/abs/1907.00990}{{\tt
  1907.00990}}].

\bibitem{osti_4444917}
M.~Demianski and E.~T. Newman, \emph{Combined kerr-nut solution of the einstein
  field equations.}, {\emph{Bull. Acad. Pol. Sci., Ser. Sci. Math. Astron.
  Phys., 14: 653- 7(1966)} }.

\bibitem{LyndenBell:1996xj}
D.~Lynden-Bell and M.~Nouri-Zonoz, \emph{{Classical monopoles: Newton, NUT
  space, gravimagnetic lensing and atomic spectra}},
  \href{http://dx.doi.org/10.1103/RevModPhys.70.427}{\emph{Rev. Mod. Phys.}
  {\bf 70} (1998) 427--446}, [\href{http://arxiv.org/abs/gr-qc/9612049}{{\tt
  gr-qc/9612049}}].

\bibitem{NouriZonoz:1998va}
M.~Nouri-Zonoz and D.~Lynden-Bell, \emph{{Gravomagnetic lensing by NUT space}},
  {\emph{Mon. Not. Roy. Astron. Soc.} {\bf 292} (1997) 714--722},
  [\href{http://arxiv.org/abs/gr-qc/9812094}{{\tt gr-qc/9812094}}].

\bibitem{Wei:2011nj}
S.-W. Wei, Y.-X. Liu, C.-E. Fu and K.~Yang, \emph{{Strong field limit analysis
  of gravitational lensing in Kerr-Taub-NUT spacetime}},
  \href{http://dx.doi.org/10.1088/1475-7516/2012/10/053}{\emph{JCAP} {\bf 10}
  (2012) 053}, [\href{http://arxiv.org/abs/1104.0776}{{\tt 1104.0776}}].

\bibitem{Rahvar:2002es}
S.~Rahvar and M.~Nouri-Zonoz, \emph{{Gravitational microlensing in NUT space}},
  \href{http://dx.doi.org/10.1046/j.1365-8711.2003.06137.x}{\emph{Mon. Not.
  Roy. Astron. Soc.} {\bf 338} (2003) 926},
  [\href{http://arxiv.org/abs/astro-ph/0204282}{{\tt astro-ph/0204282}}].

\bibitem{Rahvar:2003fh}
S.~Rahvar and F.~Habibi, \emph{{Possibility of magnetic mass detection by the
  next generation of microlensing experiments}},
  \href{http://dx.doi.org/10.1086/421728}{\emph{Astrophys. J.} {\bf 610} (2004)
  673}, [\href{http://arxiv.org/abs/astro-ph/0311242}{{\tt astro-ph/0311242}}].

\bibitem{Shen:2003pi}
J.~Q. Shen, \emph{{Gravitational analogues, geometric effects and
  gravitomagnetic charge}},
  \href{http://dx.doi.org/10.1023/A:1020082903104}{\emph{Gen. Rel. Grav.} {\bf
  34} (2002) 1423--1435}, [\href{http://arxiv.org/abs/gr-qc/0301067}{{\tt
  gr-qc/0301067}}].

\bibitem{Liu:2010ja}
C.~Liu, S.~Chen, C.~Ding and J.~Jing, \emph{{Particle Acceleration on the
  Background of the Kerr-Taub-NUT Spacetime}},
  \href{http://dx.doi.org/10.1016/j.physletb.2011.05.070}{\emph{Phys. Lett. B}
  {\bf 701} (2011) 285--290}, [\href{http://arxiv.org/abs/1012.5126}{{\tt
  1012.5126}}].

\bibitem{Pradhan:2014zia}
P.~Pradhan, \emph{{Circular geodesics in the Kerr--Newman--Taub--NUT
  spacetime}},
  \href{http://dx.doi.org/10.1088/0264-9381/32/16/165001}{\emph{Class. Quant.
  Grav.} {\bf 32} (2015) 165001}, [\href{http://arxiv.org/abs/1402.0089}{{\tt
  1402.0089}}].

\bibitem{Long:2018tij}
F.~Long, S.~Chen, J.~Wang and J.~Jing, \emph{{Electromagnetic emissions from
  near-horizon region of an extreme Kerr-Taub-Nut black hole}},
  \href{http://dx.doi.org/10.1140/epjc/s10052-019-6989-8}{\emph{Eur. Phys. J.
  C} {\bf 79} (2019) 466}, [\href{http://arxiv.org/abs/1812.11463}{{\tt
  1812.11463}}].

\bibitem{Kagramanova:2010bk}
V.~Kagramanova, J.~Kunz, E.~Hackmann and C.~Lammerzahl, \emph{{Analytic
  treatment of complete and incomplete geodesics in Taub-NUT space-times}},
  \href{http://dx.doi.org/10.1103/PhysRevD.81.124044}{\emph{Phys. Rev.} {\bf
  D81} (2010) 124044}, [\href{http://arxiv.org/abs/1002.4342}{{\tt
  1002.4342}}].

\bibitem{GarciaReyes:2004qn}
G.~Garcia-Reyes and G.~A. Gonzalez, \emph{{Charged perfect fluid disks as
  sources of Taub-NUT-type spacetimes}},
  \href{http://dx.doi.org/10.1103/PhysRevD.70.104005}{\emph{Phys. Rev. D} {\bf
  70} (2004) 104005}, [\href{http://arxiv.org/abs/0810.2575}{{\tt 0810.2575}}].

\bibitem{Chakraborty:2017nfu}
C.~Chakraborty and S.~Bhattacharyya, \emph{{Does the gravitomagnetic monopole
  exist? A clue from a black hole x-ray binary}},
  \href{http://dx.doi.org/10.1103/PhysRevD.98.043021}{\emph{Phys. Rev. D} {\bf
  98} (2018) 043021}, [\href{http://arxiv.org/abs/1712.01156}{{\tt
  1712.01156}}].

\bibitem{Bordo:2019rhu}
A.~Ballon~Bordo, F.~Gray, R.~A. Hennigar and D.~Kubiz{\v n}ák, \emph{{The
  First Law for Rotating NUTs}},
  \href{http://dx.doi.org/10.1016/j.physletb.2019.134972}{\emph{Phys. Lett. B}
  {\bf 798} (2019) 134972}, [\href{http://arxiv.org/abs/1905.06350}{{\tt
  1905.06350}}].

\bibitem{Durka:2019ajz}
R.~Durka, \emph{{The first law of black hole thermodynamics for Taub-NUT
  spacetime}},  \href{http://arxiv.org/abs/1908.04238}{{\tt 1908.04238}}.

\bibitem{Kubiznak:2019yiu}
R.~A. Hennigar, D.~Kubiz{\v n}ák and R.~B. Mann, \emph{{Thermodynamics of
  Lorentzian Taub-NUT spacetimes}},
  \href{http://dx.doi.org/10.1103/PhysRevD.100.064055}{\emph{Phys. Rev. D} {\bf
  100} (2019) 064055}, [\href{http://arxiv.org/abs/1903.08668}{{\tt
  1903.08668}}].

\bibitem{Bordo:2020kxm}
A.~Ballon~Bordo, F.~Gray and D.~Kubiz{\v n}ák, \emph{{Thermodynamics of
  Rotating NUTty Dyons}},
  \href{http://dx.doi.org/10.1007/JHEP05(2020)084}{\emph{JHEP} {\bf 05} (2020)
  084}, [\href{http://arxiv.org/abs/2003.02268}{{\tt 2003.02268}}].

\bibitem{Awad:2020dhy}
A.~Awad and S.~Eissa, \emph{{Topological dyonic Taub-Bolt/NUT-AdS solutions:
  Thermodynamics and first law}},
  \href{http://dx.doi.org/10.1103/PhysRevD.101.124011}{\emph{Phys. Rev. D} {\bf
  101} (2020) 124011}, [\href{http://arxiv.org/abs/2007.10489}{{\tt
  2007.10489}}].

\bibitem{Gibbons:1979xm}
G.~Gibbons and S.~Hawking, \emph{{Classification of Gravitational Instanton
  Symmetries}}, \href{http://dx.doi.org/10.1007/BF01197189}{\emph{Commun. Math.
  Phys.} {\bf 66} (1979) 291--310}.

\bibitem{Moynihan:2020gxj}
N.~Moynihan and J.~Murugan, \emph{{On-Shell Electric-Magnetic Duality and the
  Dual Graviton}},  \href{http://arxiv.org/abs/2002.11085}{{\tt 2002.11085}}.

\bibitem{Montonen:1977sn}
C.~Montonen and D.~I. Olive, \emph{{Magnetic Monopoles as Gauge Particles?}},
  \href{http://dx.doi.org/10.1016/0370-2693(77)90076-4}{\emph{Phys. Lett. B}
  {\bf 72} (1977) 117--120}.

\bibitem{Seiberg:1994rs}
N.~Seiberg and E.~Witten, \emph{{Electric - magnetic duality, monopole
  condensation, and confinement in N=2 supersymmetric Yang-Mills theory}},
  \href{http://dx.doi.org/10.1016/0550-3213(94)90124-4}{\emph{Nucl. Phys. B}
  {\bf 426} (1994) 19--52}, [\href{http://arxiv.org/abs/hep-th/9407087}{{\tt
  hep-th/9407087}}].

\bibitem{Seiberg:1994aj}
N.~Seiberg and E.~Witten, \emph{{Monopoles, duality and chiral symmetry
  breaking in N=2 supersymmetric QCD}},
  \href{http://dx.doi.org/10.1016/0550-3213(94)90214-3}{\emph{Nucl. Phys. B}
  {\bf 431} (1994) 484--550}, [\href{http://arxiv.org/abs/hep-th/9408099}{{\tt
  hep-th/9408099}}].

\bibitem{Olive:1995sw}
D.~I. Olive, \emph{{Exact electromagnetic duality}},
  \href{http://dx.doi.org/10.1016/0920-5632(96)00002-3}{\emph{Nucl. Phys. B
  Proc. Suppl.} {\bf 46} (1996) 1--15},
  [\href{http://arxiv.org/abs/hep-th/9508089}{{\tt hep-th/9508089}}].

\bibitem{Harvey:1996ur}
J.~A. Harvey, \emph{{Magnetic monopoles, duality and supersymmetry}},  in
  \emph{{ICTP Summer School in High-energy Physics and Cosmology}}, 3, 1996.
\newblock \href{http://arxiv.org/abs/hep-th/9603086}{{\tt hep-th/9603086}}.

\bibitem{Olive:1997fg}
D.~I. Olive, \emph{{Introduction to electromagnetic duality}},
  \href{http://dx.doi.org/10.1016/S0920-5632(97)00412-X}{\emph{Nucl. Phys. B
  Proc. Suppl.} {\bf 58} (1997) 43--55}.

\bibitem{Kawai:1985xq}
H.~Kawai, D.~Lewellen and S.~Tye, \emph{{A Relation Between Tree Amplitudes of
  Closed and Open Strings}},
  \href{http://dx.doi.org/10.1016/0550-3213(86)90362-7}{\emph{Nucl. Phys. B}
  {\bf 269} (1986) 1--23}.

\bibitem{Bern:2008qj}
Z.~Bern, J.~Carrasco and H.~Johansson, \emph{{New Relations for Gauge-Theory
  Amplitudes}}, \href{http://dx.doi.org/10.1103/PhysRevD.78.085011}{\emph{Phys.
  Rev. D} {\bf 78} (2008) 085011}, [\href{http://arxiv.org/abs/0805.3993}{{\tt
  0805.3993}}].

\bibitem{Bern:2010ue}
Z.~Bern, J.~J.~M. Carrasco and H.~Johansson, \emph{{Perturbative Quantum
  Gravity as a Double Copy of Gauge Theory}},
  \href{http://dx.doi.org/10.1103/PhysRevLett.105.061602}{\emph{Phys. Rev.
  Lett.} {\bf 105} (2010) 061602}, [\href{http://arxiv.org/abs/1004.0476}{{\tt
  1004.0476}}].

\bibitem{Bern:2010yg}
Z.~Bern, T.~Dennen, Y.-t. Huang and M.~Kiermaier, \emph{{Gravity as the Square
  of Gauge Theory}},
  \href{http://dx.doi.org/10.1103/PhysRevD.82.065003}{\emph{Phys. Rev. D} {\bf
  82} (2010) 065003}, [\href{http://arxiv.org/abs/1004.0693}{{\tt 1004.0693}}].

\bibitem{Monteiro:2014cda}
R.~Monteiro, D.~O'Connell and C.~D. White, \emph{{Black holes and the double
  copy}}, \href{http://dx.doi.org/10.1007/JHEP12(2014)056}{\emph{JHEP} {\bf 12}
  (2014) 056}, [\href{http://arxiv.org/abs/1410.0239}{{\tt 1410.0239}}].

\bibitem{Luna:2015paa}
A.~Luna, R.~Monteiro, D.~O'Connell and C.~D. White, \emph{{The classical double
  copy for Taub--NUT spacetime}},
  \href{http://dx.doi.org/10.1016/j.physletb.2015.09.021}{\emph{Phys. Lett. B}
  {\bf 750} (2015) 272--277}, [\href{http://arxiv.org/abs/1507.01869}{{\tt
  1507.01869}}].

\bibitem{Banerjee:2019saj}
A.~Banerjee, E.~Colg\'ain, J.~Rosabal and H.~Yavartanoo, \emph{{Ehlers as EM
  duality in the double copy}},  \href{http://arxiv.org/abs/1912.02597}{{\tt
  1912.02597}}.

\bibitem{Bahjat-Abbas:2020cyb}
N.~Bahjat-Abbas, R.~Stark-Much\~ao and C.~D. White, \emph{{Monopoles,
  shockwaves and the classical double copy}},
  \href{http://dx.doi.org/10.1007/JHEP04(2020)102}{\emph{JHEP} {\bf 04} (2020)
  102}, [\href{http://arxiv.org/abs/2001.09918}{{\tt 2001.09918}}].

\bibitem{Ridgway:2015fdl}
A.~K. Ridgway and M.~B. Wise, \emph{{Static Spherically Symmetric Kerr-Schild
  Metrics and Implications for the Classical Double Copy}},
  \href{http://dx.doi.org/10.1103/PhysRevD.94.044023}{\emph{Phys. Rev. D} {\bf
  94} (2016) 044023}, [\href{http://arxiv.org/abs/1512.02243}{{\tt
  1512.02243}}].

\bibitem{Carrillo-Gonzalez:2017iyj}
M.~Carrillo-Gonz\'alez, R.~Penco and M.~Trodden, \emph{{The classical double
  copy in maximally symmetric spacetimes}},
  \href{http://dx.doi.org/10.1007/JHEP04(2018)028}{\emph{JHEP} {\bf 04} (2018)
  028}, [\href{http://arxiv.org/abs/1711.01296}{{\tt 1711.01296}}].

\bibitem{Goldberger:2017vcg}
W.~D. Goldberger and A.~K. Ridgway, \emph{{Bound states and the classical
  double copy}},
  \href{http://dx.doi.org/10.1103/PhysRevD.97.085019}{\emph{Phys. Rev. D} {\bf
  97} (2018) 085019}, [\href{http://arxiv.org/abs/1711.09493}{{\tt
  1711.09493}}].

\bibitem{Bahjat-Abbas:2017htu}
N.~Bahjat-Abbas, A.~Luna and C.~D. White, \emph{{The Kerr-Schild double copy in
  curved spacetime}},
  \href{http://dx.doi.org/10.1007/JHEP12(2017)004}{\emph{JHEP} {\bf 12} (2017)
  004}, [\href{http://arxiv.org/abs/1710.01953}{{\tt 1710.01953}}].

\bibitem{Lee:2018gxc}
K.~Lee, \emph{{Kerr-Schild Double Field Theory and Classical Double Copy}},
  \href{http://dx.doi.org/10.1007/JHEP10(2018)027}{\emph{JHEP} {\bf 10} (2018)
  027}, [\href{http://arxiv.org/abs/1807.08443}{{\tt 1807.08443}}].

\bibitem{Kim:2019jwm}
K.~Kim, K.~Lee, R.~Monteiro, I.~Nicholson and D.~Peinador~Veiga, \emph{{The
  Classical Double Copy of a Point Charge}},
  \href{http://dx.doi.org/10.1007/JHEP02(2020)046}{\emph{JHEP} {\bf 02} (2020)
  046}, [\href{http://arxiv.org/abs/1912.02177}{{\tt 1912.02177}}].

\bibitem{delaCruz:2020bbn}
L.~de~la Cruz, B.~Maybee, D.~O'Connell and A.~Ross, \emph{{Classical Yang-Mills
  observables from amplitudes}},
  \href{http://dx.doi.org/10.1007/JHEP12(2020)076}{\emph{JHEP} {\bf 12} (2020)
  076}, [\href{http://arxiv.org/abs/2009.03842}{{\tt 2009.03842}}].

\bibitem{Arkani-Hamed:2017jhn}
N.~Arkani-Hamed, T.-C. Huang and Y.-t. Huang, \emph{{Scattering Amplitudes For
  All Masses and Spins}},  \href{http://arxiv.org/abs/1709.04891}{{\tt
  1709.04891}}.

\bibitem{Weinberg:1965rz}
S.~Weinberg, \emph{{Photons and gravitons in perturbation theory: Derivation of
  Maxwell's and Einstein's equations}},
  \href{http://dx.doi.org/10.1103/PhysRev.138.B988}{\emph{Phys. Rev.} {\bf 138}
  (1965) B988--B1002}.

\bibitem{Kol:2020ucd}
U.~Kol and M.~Porrati, \emph{{Gravitational Wu-Yang Monopoles}},
  \href{http://arxiv.org/abs/2003.09054}{{\tt 2003.09054}}.

\bibitem{Visinescu:1999zs}
M.~Visinescu, \emph{{Generalized Taub - NUT metrics and Killing-Yano tensors}},
  \href{http://dx.doi.org/10.1088/0305-4470/33/23/312}{\emph{J. Phys. A} {\bf
  33} (2000) 4383--4392}, [\href{http://arxiv.org/abs/hep-th/9911126}{{\tt
  hep-th/9911126}}].

\bibitem{Bern:1994zx}
Z.~Bern, L.~J. Dixon, D.~C. Dunbar and D.~A. Kosower, \emph{{One loop n point
  gauge theory amplitudes, unitarity and collinear limits}},
  \href{http://dx.doi.org/10.1016/0550-3213(94)90179-1}{\emph{Nucl. Phys. B}
  {\bf 425} (1994) 217--260}, [\href{http://arxiv.org/abs/hep-ph/9403226}{{\tt
  hep-ph/9403226}}].

\bibitem{Bern:1994cg}
Z.~Bern, L.~J. Dixon, D.~C. Dunbar and D.~A. Kosower, \emph{{Fusing gauge
  theory tree amplitudes into loop amplitudes}},
  \href{http://dx.doi.org/10.1016/0550-3213(94)00488-Z}{\emph{Nucl. Phys. B}
  {\bf 435} (1995) 59--101}, [\href{http://arxiv.org/abs/hep-ph/9409265}{{\tt
  hep-ph/9409265}}].

\bibitem{Holstein:2004dn}
B.~R. Holstein and J.~F. Donoghue, \emph{{Classical physics and quantum
  loops}}, \href{http://dx.doi.org/10.1103/PhysRevLett.93.201602}{\emph{Phys.
  Rev. Lett.} {\bf 93} (2004) 201602},
  [\href{http://arxiv.org/abs/hep-th/0405239}{{\tt hep-th/0405239}}].

\bibitem{Neill:2013wsa}
D.~Neill and I.~Z. Rothstein, \emph{{Classical Space-Times from the S Matrix}},
  \href{http://dx.doi.org/10.1016/j.nuclphysb.2013.09.007}{\emph{Nucl. Phys. B}
  {\bf 877} (2013) 177--189}, [\href{http://arxiv.org/abs/1304.7263}{{\tt
  1304.7263}}].

\bibitem{Guevara:2017csg}
A.~Guevara, \emph{{Holomorphic Classical Limit for Spin Effects in
  Gravitational and Electromagnetic Scattering}},
  \href{http://dx.doi.org/10.1007/JHEP04(2019)033}{\emph{JHEP} {\bf 04} (2019)
  033}, [\href{http://arxiv.org/abs/1706.02314}{{\tt 1706.02314}}].

\bibitem{Bern:2019crd}
Z.~Bern, C.~Cheung, R.~Roiban, C.-H. Shen, M.~P. Solon and M.~Zeng,
  \emph{{Black Hole Binary Dynamics from the Double Copy and Effective
  Theory}}, \href{http://dx.doi.org/10.1007/JHEP10(2019)206}{\emph{JHEP} {\bf
  10} (2019) 206}, [\href{http://arxiv.org/abs/1908.01493}{{\tt 1908.01493}}].

\bibitem{Chung:2018kqs}
M.-Z. Chung, Y.-T. Huang, J.-W. Kim and S.~Lee, \emph{{The simplest massive
  S-matrix: from minimal coupling to Black Holes}},
  \href{http://dx.doi.org/10.1007/JHEP04(2019)156}{\emph{JHEP} {\bf 04} (2019)
  156}, [\href{http://arxiv.org/abs/1812.08752}{{\tt 1812.08752}}].

\bibitem{Chung:2019duq}
M.-Z. Chung, Y.-T. Huang and J.-W. Kim, \emph{{Classical potential for general
  spinning bodies}},
  \href{http://dx.doi.org/10.1007/JHEP09(2020)074}{\emph{JHEP} {\bf 09} (2020)
  074}, [\href{http://arxiv.org/abs/1908.08463}{{\tt 1908.08463}}].

\bibitem{Chung:2019yfs}
M.-Z. Chung, Y.-T. Huang and J.-W. Kim, \emph{{Kerr-Newman stress-tensor from
  minimal coupling to all orders in spin}},
  \href{http://arxiv.org/abs/1911.12775}{{\tt 1911.12775}}.

\bibitem{Chung:2020rrz}
M.-Z. Chung, Y.-t. Huang, J.-W. Kim and S.~Lee, \emph{{Complete Hamiltonian for
  spinning binary systems at first post-Minkowskian order}},
  \href{http://arxiv.org/abs/2003.06600}{{\tt 2003.06600}}.

\bibitem{Guevara:2018wpp}
A.~Guevara, A.~Ochirov and J.~Vines, \emph{{Scattering of Spinning Black Holes
  from Exponentiated Soft Factors}},
  \href{http://dx.doi.org/10.1007/JHEP09(2019)056}{\emph{JHEP} {\bf 09} (2019)
  056}, [\href{http://arxiv.org/abs/1812.06895}{{\tt 1812.06895}}].

\bibitem{Guevara:2019fsj}
A.~Guevara, A.~Ochirov and J.~Vines, \emph{{Black-hole scattering with general
  spin directions from minimal-coupling amplitudes}},
  \href{http://dx.doi.org/10.1103/PhysRevD.100.104024}{\emph{Phys. Rev. D} {\bf
  100} (2019) 104024}, [\href{http://arxiv.org/abs/1906.10071}{{\tt
  1906.10071}}].

\bibitem{Moynihan:2019bor}
N.~Moynihan, \emph{{Kerr-Newman from Minimal Coupling}},
  \href{http://dx.doi.org/10.1007/JHEP01(2020)014}{\emph{JHEP} {\bf 01} (2020)
  014}, [\href{http://arxiv.org/abs/1909.05217}{{\tt 1909.05217}}].

\bibitem{Forde:2007mi}
D.~Forde, \emph{{Direct extraction of one-loop integral coefficients}},
  \href{http://dx.doi.org/10.1103/PhysRevD.75.125019}{\emph{Phys. Rev. D} {\bf
  75} (2007) 125019}, [\href{http://arxiv.org/abs/0704.1835}{{\tt 0704.1835}}].

\bibitem{Kosower:2018adc}
D.~A. Kosower, B.~Maybee and D.~O'Connell, \emph{{Amplitudes, Observables, and
  Classical Scattering}},
  \href{http://dx.doi.org/10.1007/JHEP02(2019)137}{\emph{JHEP} {\bf 02} (2019)
  137}, [\href{http://arxiv.org/abs/1811.10950}{{\tt 1811.10950}}].

\bibitem{Caron-Huot:2018ape}
S.~Caron-Huot and Z.~Zahraee, \emph{{Integrability of Black Hole Orbits in
  Maximal Supergravity}},
  \href{http://dx.doi.org/10.1007/JHEP07(2019)179}{\emph{JHEP} {\bf 07} (2019)
  179}, [\href{http://arxiv.org/abs/1810.04694}{{\tt 1810.04694}}].

\bibitem{Johansson:2019dnu}
H.~Johansson and A.~Ochirov, \emph{{Double copy for massive quantum particles
  with spin}}, \href{http://dx.doi.org/10.1007/JHEP09(2019)040}{\emph{JHEP}
  {\bf 09} (2019) 040}, [\href{http://arxiv.org/abs/1906.12292}{{\tt
  1906.12292}}].

\bibitem{Cheung:2020sdj}
C.~Cheung and M.~P. Solon, \emph{{Tidal Effects in the Post-Minkowskian
  Expansion}},  \href{http://arxiv.org/abs/2006.06665}{{\tt 2006.06665}}.

\bibitem{Brustein:2020tpg}
R.~Brustein and Y.~Sherf, \emph{{Quantum Love}},
  \href{http://arxiv.org/abs/2008.02738}{{\tt 2008.02738}}.

\bibitem{Csaki:2020inw}
C.~Csaki, S.~Hong, Y.~Shirman, O.~Telem, J.~Terning and M.~Waterbury,
  \emph{{Scattering Amplitudes for Monopoles: Pairwise Little Group and
  Pairwise Helicity}},  \href{http://arxiv.org/abs/2009.14213}{{\tt
  2009.14213}}.

\bibitem{Porto:2016zng}
R.~A. Porto, \emph{{The Tune of Love and the Nature(ness) of Spacetime}},
  \href{http://dx.doi.org/10.1002/prop.201600064}{\emph{Fortsch. Phys.} {\bf
  64} (2016) 723--729}, [\href{http://arxiv.org/abs/1606.08895}{{\tt
  1606.08895}}].

\bibitem{Kim:2020dif}
J.-W. Kim and M.~Shim, \emph{{Sum rule for Love}},
  \href{http://arxiv.org/abs/2011.03337}{{\tt 2011.03337}}.

\bibitem{Hagen:1965zz}
C.~Hagen, \emph{{Noncovariance of the Dirac Monopole}},
  \href{http://dx.doi.org/10.1103/PhysRev.140.B804}{\emph{Phys. Rev.} {\bf 140}
  (1965) B804--B810}.

\bibitem{Zwanziger:1970hk}
D.~Zwanziger, \emph{{Local Lagrangian quantum field theory of electric and
  magnetic charges}},
  \href{http://dx.doi.org/10.1103/PhysRevD.3.880}{\emph{Phys. Rev. D} {\bf 3}
  (1971) 880}.

\bibitem{Schwarz:1993vs}
J.~H. Schwarz and A.~Sen, \emph{{Duality symmetric actions}},
  \href{http://dx.doi.org/10.1016/0550-3213(94)90053-1}{\emph{Nucl. Phys. B}
  {\bf 411} (1994) 35--63}, [\href{http://arxiv.org/abs/hep-th/9304154}{{\tt
  hep-th/9304154}}].

\bibitem{Dirac:1948um}
P.~A. Dirac, \emph{{The Theory of magnetic poles}},
  \href{http://dx.doi.org/10.1103/PhysRev.74.817}{\emph{Phys. Rev.} {\bf 74}
  (1948) 817--830}.

\bibitem{Schwinger:1966nj}
J.~S. Schwinger, \emph{{Magnetic charge and quantum field theory}},
  \href{http://dx.doi.org/10.1103/PhysRev.144.1087}{\emph{Phys. Rev.} {\bf 144}
  (1966) 1087--1093}.

\bibitem{ArkaniHamed:2006dz}
N.~Arkani-Hamed, L.~Motl, A.~Nicolis and C.~Vafa, \emph{{The String landscape,
  black holes and gravity as the weakest force}},
  \href{http://dx.doi.org/10.1088/1126-6708/2007/06/060}{\emph{JHEP} {\bf 06}
  (2007) 060}, [\href{http://arxiv.org/abs/hep-th/0601001}{{\tt
  hep-th/0601001}}].

\bibitem{Buonanno:1998gg}
A.~Buonanno and T.~Damour, \emph{{Effective one-body approach to general
  relativistic two-body dynamics}},
  \href{http://dx.doi.org/10.1103/PhysRevD.59.084006}{\emph{Phys. Rev. D} {\bf
  59} (1999) 084006}, [\href{http://arxiv.org/abs/gr-qc/9811091}{{\tt
  gr-qc/9811091}}].

\bibitem{Damour:2016gwp}
T.~Damour, \emph{{Gravitational scattering, post-Minkowskian approximation and
  Effective One-Body theory}},
  \href{http://dx.doi.org/10.1103/PhysRevD.94.104015}{\emph{Phys. Rev. D} {\bf
  94} (2016) 104015}, [\href{http://arxiv.org/abs/1609.00354}{{\tt
  1609.00354}}].

\end{thebibliography}\endgroup
\bibliographystyle{JHEP}

\end{document}